\documentclass[nofootinbib,amsmath,prd,aps,superscriptaddress,tightenlines,preprint,preprintnumbers,letterpaper,12pt]{revtex4}

\usepackage{color}
\definecolor{lcolor}{rgb}{0,0.2,0.6}
\usepackage[bookmarks=false,
	colorlinks=true,
	linkcolor = black,
	urlcolor = lcolor,
	citecolor = lcolor]{hyperref}            
\usepackage{graphicx}
\usepackage{epstopdf}
\usepackage{bm}
\usepackage{epsfig}
\usepackage{graphics}
\usepackage{xspace}
\usepackage{slashed}
\usepackage{feynmp}
\usepackage{ams fonts}
\usepackage{sidecap}
\usepackage{dcolumn} % Align table columns on decimal point
\usepackage{longtable} % long tables
\usepackage{booktabs}

\hypersetup{pdfpagemode=FullScreen}

\def\OMIT#1{}

\newcommand{\nn}{\nonumber}

\newcommand{\bea}{\begin{eqnarray}}
\newcommand{\eea}{\end{eqnarray}}

\newcommand{\fslash}{ f \!\!\!\slash}
\newcommand{\Deltaslash}{ \Delta \!\!\!\!\slash}

\newcommand{\sld}[1]{\slashed #1}

\newcommand{\ep}[4]{\epsilon^{#1#2#3#4}}

\begin{document}
\setlength\baselineskip{17pt}

%%%%%%%%%%%%%%%%%%%%%%%%%%%%%%%%%%%%%%%%%%
%Define Title, Author, Address, Preprint#

\preprint{\vbox{ \hbox{NPAC-12-07} }}

\title{Higher Twist in Electroproduction: Flavor Non-Singlet QCD Evolution}
\vspace{0.3 in}

  \author{Michael J. Glatzmaier\footnote{Electronic address: \href{mailto:michael.glatzmaier@gmail.com}{michael.glatzmaier@gmail.com}}}\affiliation{Department of Physics and Astronomy, University of Kentucky, Lexington, KY 40506}

\author{Sonny Mantry\footnote{Electronic address:  \href{mailto:mantry147@gmail.com}{mantry147@gmail.com}}}
 \affiliation{High Energy Division, Argonne National Laboratory,
Argonne, IL 60439}\affiliation{Department of Physics and Astronomy, Northwestern
University, Evanston, IL 60208}
 
\author{Michael J. Ramsey-Musolf\footnote{Electronic address:  \href{mailto:mjrm@physics.wisc.edu}{mjrm@physics.wisc.edu}}}
\affiliation{University of Madison-Wisconsin,Madison,WI,53706}

\date{\today\\ \vspace{1cm} }

%%%%%%%%%%%%%%%%%%%%%%%%%%%%%%%%%%%%%%%%%%
%Create the title page

\begin{abstract}
%\vspace*{0.3cm}

We present results for the one loop anomalous dimension matrix of flavor non-singlet twist-4 operators of lowest spin that contribute to the leading moment of the $F_2$ structure function in deep inelastic electron-nucleon scattering.  We analyze the flavor structure of the anomalous dimension matrix and decompose the leading moment of $F_2$ into separate flavor channels.  In addition to building on previous work with higher-twist operators, these results can provide a benchmark for future work that generalizes to include the higher moments as well. We include non-perturbative input from the lattice and phenomenological estimates of the twist-4 matrix elements and estimate the twist-4 contributions to the leading moment of $F_2$.  The results suggest that the overall twist-4 contribution may be suppressed due to either cancellations among the twist-4 terms or inherently small twist-4 matrix elements.

\end{abstract}
\maketitle

%%%%%%%%%%%%%%%%%%%%%%%%%%%%%%%%%%%%%%%%%%
\newpage
%%%%%%%%%%%%%%%%%%%%%%%%%%%%%%%%%%%%%%%%%%
%Main body of the paper

\section{Introduction}\label{intro}

In this paper we report on a computation of the leading log $Q^2$ evolution of the twist-4 contribution to the flavor non-singlet, leading moment of the $F_2$ structure function and provide phenomenological estimates of the resulting effects. Our study is motivated broadly by one of the major challenges for nuclear physics:
understanding the dynamics of quarks and gluons, as determined by the QCD Lagrangian, and explaining their connection to the hadronic degrees of freedom.  The twist expansion in QCD is a useful tool that is well-suited for this challenge.  The property of asymptotic freedom of QCD allows one to calculate sufficiently inclusive hadronic observables, at asymptotically high energies, in terms of the perturbative quark and gluon degrees of freedom.  This phenomenon is often referred to as quark-hadron duality.  One of the simplest examples of this duality is  the process $e^+e^-\to \> hadrons$ which is described well by the quark-level process $e^+e^-\to q\bar{q}\>$ away from thresholds. Similarly,  deep inelastic electron-nucleon scattering is described by electrons scattering off free quarks in the asymptotic region.  

However at low momentum scales, where the strong coupling $\alpha_s$ is large, multi-parton correlations become important and lead to violations of quark-hadron duality.  These correlations are embodied in higher twist effects.  At sufficiently low scales, QCD is non-perturbative and the hadronic bound states of quarks and gluons become the relevant degrees of freedom.  Despite this, there are several examples where low energy hadronic observables, averaged over appropriate intervals, exhibit behavior that reveal the underlying connection to the quark and gluon degrees of freedom.   Bloom and Gilman ~\cite{Bloom:1970xb,Bloom:1971ye,Stein:1975yy} first observed that electron-nucleon scattering in the resonance region is related to the deep inelastic scaling regime. In particular, they observed that even in the region of low momentum transfer ($Q^2$), dominated by the highly non-perturbative dynamics of nucleon resonances, the nucleon structure function $F_2(x,Q^2)$ exhibits logarithmic scaling in $Q^2$ when averaged over appropriate intervals in Bjorken-$x$. Furthermore, the resonance structures seen in $F_2(x,Q^2)$ as a function of the Bjorken-$x$, slide along the deep inelastic scaling curve for increasing $Q^2$. This logarithmic scaling in $Q^2$ of the structure function $F_2$ is described by  the DGLAP evolution of the leading twist parton distribution functions (PDFs). This manifestation of quark-hadron duality which relates the resonance region to the deep inelastic scaling region is known as the Bloom-Gilman (BG) duality.  At low values of $Q^2$, one expects the onset of power law behavior corresponding to contributions from higher twist terms in the operator product expansion (OPE). Such behavior would signal a clear violation of BG duality and give a direct probe of multi-parton correlations in the nucleon.

Detailed studies of the BG duality and its violation can provide insight into the dynamics of the quark-hadron transition and have lead to a large experimental effort.  Since the early days of the SLAC-MIT~\cite{Stein:1975yy} experiment, a wealth of data  (for a comprehensive review see Ref.\cite{Melnitchouk:2005zr})  on  structure functions has been accumulated over a wide range in $x$ and $Q^2$. A large fraction of this data~\cite{Bosted:1993cc, Chekanov:2001qu, Niculescu:1999mw, Ricco:1997zm, Niculescu:2005rh, Armstrong:2001xj,Osipenko:2003bu}   is on the proton  structure function $F_2^p(x,Q^2)$ which is now the best measured quantity in deep inelastic electron scattering. Data is also available on deuterium~\cite{Osipenko:2010sb} and heavy nuclear~\cite{Osipenko:2006gu}  targets in the high-$x$ and low $Q^2$ region. The data on the structure functions in the resonance region has been compared to the scaling curves obtained from global fits~\cite{Lai:1999wy, Martin:1998np} of the PDFs and DGLAP evolution.  The resonance peak structures seen in the $F_2^p(x,Q^2)$ structure function are observed on average to oscillate around the scaling curve. In particular, the average of the structure function over all values of $x$, including over all resonance peaks, exhibits scaling behavior. Furthermore,   the average of $F_2^p(x,Q^2)$ over individual resonance peaks is also observed to follow the scaling curve and is known as \textit{local} BG duality. Similar scaling behavior is also observed for deuterium and heavy nuclei targets.

In modern field-theoretic language, quark-hadron duality can be quantitatively formulated~\cite{DeRujula:1976ke, DeRujula:1976tz} in terms of  the operator product expansion (OPE).  The Cornwall-Norton moments of the $F_2$ nucleon structure function 
\bea
M_2^{(n)}(Q^2) = \int_0^1 dx \>x^{n-2} F_2(x,Q^2),
\eea
 can be expressed schematically in terms of the OPE as
\begin{equation}\label{qh2}
M_2^{(n)}(Q^2) =  \sum_i\sum_{\tau=2k}^{\infty}\left(\frac{\Lambda^2}{Q^2}\right)^{\frac{\tau-2}{2}} C_{n \tau}^i (\mu,Q^2)\ \ \frac{\langle \mathcal{O}_{n\tau}^i\rangle}{\Lambda^{\tau-2}} .
\end{equation}   
Here $k$ runs over all positive integers and the indices $n,\tau, i$ denote the spin, twist, and type of operators respectively. The twist is defined $\tau=d-s$, where $d,s$ denote the dimension and spin of the operator $\mathcal{O}_{n\tau}^i$. The Wilson coefficients $C_{n\tau}^i$ are perturbatively calculable as an expansion in $\alpha_s(Q^2)$ and exhibit logarithmic scaling in $Q^2/\mu^2$ with $\mu$ being an appropriately chosen input scale. The non-perturbative nucleon matrix element of the operator $ \mathcal{O}_{n\tau}^i$ is denoted by $\langle \mathcal{O}_{n\tau}^i\rangle $ and has been scaled to an appropriate power of a typical hadronic scale $\Lambda\sim 1$ GeV. The power law behavior in $Q^2$ of the various terms in the OPE is determined by the twist $\tau$. The leading twist ($\tau=2$) nucleon matrix elements are given by the moments of the standard PDFs. Quark-hadron duality  corresponds to the dominance of the leading twist terms which are determined by the scattering of  electrons from almost free quarks weighted by the PDFs. Logarithmic corrections to Bjorken-scaling are determined by the  standard DGLAP evolution of the PDFs. Violations of quark-hadron duality arise from the higher twist terms in the OPE as power corrections in $1/Q^2$. These higher twist terms encode long range multi-parton correlations in the nucleon and are expected to become important at low $Q^2$. 

In the language of the OPE, the observed BG duality corresponds to unexpectedly small contributions from the higher twist terms to  the lowest ($n=2$) moment of the $F_2$ structure function at low $Q^2$. The higher moments of the structure function, weighted more by the large $x$ resonance region, are expected to be more sensitive to higher twist effects. A recent analysis by the CLAS collaboration \cite{Osipenko:2003bu} found that the moments of the $F_2^p(x,Q^2)$ structure function were dominated by the leading twist terms down to $Q^2 \sim 1$ GeV$^2$, implying correspondingly small higher twist effects.  For the lowest moment, after accounting for kinematic power corrections, the higher twist contributions were less than about $5\%$ of the leading twist moment for $Q^2>1\>\text{GeV}^2$.  These results were obtained through a detailed study of the $Q^2$ behavior of the collected data. 

The high quality of available data allows for a systematic study of higher twist correlations, providing a window into quark-hadron duality violations and nucleon structure.  One limitation for such a program is the lack of precise theoretical knowledge of the renormalization group (RG) evolution of the higher twist operators.  Given the absence of this theoretical input, the CLAS collaboration considered the effects of twist-4 and twist-6 contributions, in addition to the leading twist effects, using a simple {\em ansatz}~\cite{Ricco:1998yr, Simula:2000ka} that parameterizes these contributions to the moments of the structure function as
\bea
\label{param}
M_2^{(n)}(Q^2) = \eta_n (Q^2) + a_n^{(4)} \left [ \frac{\alpha_s(Q^2)}{\alpha_s(\mu^2)} \right ]^{\gamma_n^{(4)}} \frac{\mu^2}{Q^2} + a_n^{(6)} \left [ \frac{\alpha_s(Q^2)}{\alpha_s(\mu^2)} \right ]^{\gamma_n^{(6)}} \frac{\mu^4}{Q^4},
\eea
where $\eta_n(Q^2)$ is the leading twist contribution, $a_n^{(4)}$ and $a_n^{(6)}$ parameterize the twist-4 and twist-6 nucleon matrix elements respectively, and $\gamma_n^{(4)}$ and $\gamma_n^{(6)}$ are effective anomalous dimensions parametrizing the RG evolution of twist-4 and twist-6 operators respectively.  With the parameterization written in Eq.(\ref{param}), the CLAS collaboration interpreted the unexpectedly tiny higher twist contribution to the moment as being due to a conspiracy of cancellation between twist-4 and an oppositely signed twist-6 contribution. From a rigorous theoretical perspective, however, the situation is considerably more complex, as higher twist contributions are determined by a large number of operators that mix under RG evolution.  Both the contributions from these matrix elements and the details of their mixing are ignored in Eq.(\ref{param}).  A basis of operators at twist-4 along with their tree level Wilson coefficients was first given in Refs.~\cite{Jaffe:1981td, Jaffe:1981sz, Jaffe:1982pm}, in the transverse basis in Ref.~\cite{Ellis:1982cd}, and more recently using the soft-collinear effective theory (SCET) in Ref.~\cite{Marcantonini:2008qn}.  A conformal basis of higher twist operators was constructed in Ref.~\cite{Braun:2008ia} and a one-loop analysis of conformal higher twist operators is presented in Refs.~\cite{Braun:2008ia, Braun:2009vc}.  

%A complete analysis of higher twist effects must necessarily include estimates of the non-perturbative matrix elements which contribute to Eq.(\ref{qh2}) as well.  To this end, knowledge of the RG evolution of higher twist operators benefit the lattice community since knowledge of the anomalous dimension matrix reduces the number of fit parameters appearing in Eq.(\ref{param}).  This in turn allows for more accurate extractions of higher twist matrix elements from DIS data which can be used to check lattice computations. 

Higher twist operators in QCD are also of interest for parity violating deep inelastic scattering (PVDIS).  As part of the 12 GeV upgrade at JLAB, new experiments~\cite{Souder:2008zz,Subedi:2009zz} will measure the electron polarization asymmetry,  
\bea
A_{RL}=\frac{\sigma_R-\sigma_L}{\sigma_R+\sigma_L}
\eea
in parity violating deep inelastic scattering off a deuteron target over a wide range of $Q^2$ and $x$ to sub-precent level precision.  The impact of the logarithmic running of higher twist operators is  one  effect one must account for when interpreting the asymmetry.  Due to the high precision of the measurements, hadronic uncertainties including higher-twist effects must be investigated carefully as they can potentially cloud theoretical interpretations of deviations from Standard Model (SM) predictions.  The effects of higher twist contributions to this parity-violating asymmetry were recently studied in Refs.~\cite{Hobbs:2008mm, Mantry:2010ki}. Based on the argument by Bjorken~\cite{Bjorken:1978ry} and Wolfenstein ~\cite{Wolfenstein:1978rr} it was shown~\cite{Mantry:2010ki} that this asymmetry can be a powerfufl probe of quark-quark correlations in the nucleon. For a deuterium target, $A_{RL}$ is sensitive to a single four-quark operator involving up and down-quark fields
\bea\label{pvdisop}
Q^{\mu\nu}_{ud}(x)=\frac{1}{2}[\bar{u}(x)\gamma^\mu u(x)d(0)\gamma^\nu d(0)+(u\leftrightarrow d)]
\eea
which is a twist-4 operator.  As pointed out in~\cite{Mantry:2010ki}, combining high precision data taken over a wide range of $x$ and $Q^2$ from future PVDIS experiments at JLAB as well as electron ion collider (EIC) data, may allow a separation of higher twist contributions to $A_{RL}$ from the CSV effects depending on their relative sizes.  In order to so however, one must have an accurate determination of the size of the matrix element of the operator in Eq.(\ref{pvdisop}) as well as accurate knowledge of its mixing with other twist-4 operators under the renormalization group. 

 In light of these experimental developments as well as the broader goal of eludicating the transition from perturbative to non-perturbative features of nucleon structure, we have undertaken the present study.
Our goal in this work is to compute the one-loop anomalous dimension matrix for flavor non-singlet twist-4 operators at lowest spin, including all mixing effects.  Our calculations of the anomalous dimension matrix are an extension of previous works~\cite{Gottlieb:1978zj,Okawa:1980ei,Okawa:1981uw} where parts of the one-loop anomalous dimension matrix were computed. In particular, the graphs of Fig.~\ref{4qgraphs}(B)(below) were not included in previous analyses, and our results for the graphs of Fig.~\ref{4qgraphs}(A) differ from those computed in~\cite{Gottlieb:1978zj}.  We defer the calculation of the RG evolution of gluonic operators and twist-4 operators of arbitrary spin for future work. To this end, we have also listed the iso-singlet flavor operators which mix with gluon operators.    
  
We have combined our perturbative calculations with input from the lattice~\cite{Gockeler:2001xw} and phenomenological estimates~\cite{Choi:1993cu} to provide illustrative computations of the evolution of the isovector, flavor non-singlet contribution to $F_2$ in a range of $Q^2$ so as to make contact with the CLAS analysis~\cite{Osipenko:2003bu}.  We demonstrate that theoretically one expects a relatively tiny overall twist-4 contribution to the moment in the resonance region, having a magnitude that is consistent with the CLAS analysis. 
We also show that within twist-4, cancellations or enhancements can occur between different flavor channels contributing to the leading moment of $F_2$.  Thus, a suppression of the twist-4 contribution may be due either to cancellations between different operator contributions or to relatively small individual matrix elements themselves. Our key  results can be summarized in Eq.(\ref{moment}) and Figs.[\ref{octet}] and [\ref{RGE}].  These results demonstrate that probing higher twist effects in the leading moment of $F_2$ would require a substantial improvement in experimental precision. In particular, the observation of any breakdown of cancellations due to $Q^2$-evolution would likely require a substantial reduction in experimental error. As a corollary, we also note that a complete QCD analysis of twist-four contributions will require new non-perturbative computations of the twist-4 operator matrix elements, as the illustrative results given in our study have required making an {\em ansatz} about the values of several of these matrix elements.

This paper is organized as follows,  in section \ref{formalism} we review the standard formalism of the operator product expansion (OPE) and establish basic notation.  In section \ref{basis}, we review the leading twist basis of operators and list the basis of quark and gluonic twist-4 operators. In section \ref{anom-dim}, we present both the Feynman diagrams and renormalization factors for the basis of operators introduced in section \ref{basis}, and in section \ref{counting} we introduce a power counting scheme to ensure the anomalous dimension has a consistent power in the strong coupling.  In section \ref{sec-flavor} we organize the basis of twist-4 quark operators in terms of the irreducible representations of the flavor group $\text{SU(3)}_f$.  In sections \ref{wcs} we list the tree level Wilson coefficients used to plot the leading moment of $F_2$ and, in section \ref{matrixelements}, we estimate values for the twist-4 reduced matrix elements based on lattice computations and model independent estimates.  Finally in section \ref{renorm}, we present our results for the leading log evolution of $F_2(x,Q^2)$.  We discuss these results and comment on future work in section \ref{disc}.

%%%%%%%%%%%%%%%%%%%%%%%%%%%%%%%%%%%%%%%
%												                            %
%				GENERAL FORMALISM				                            %
%											                                     %
%%%%%%%%%%%%%%%%%%%%%%%%%%%%%%%%%%%%%%%
\section{General Formalism}\label{formalism}
In this section we review the formalism and relevant notation for electron-nucleon deep inelastic scattering (DIS). The differential cross-section in the one photon exchange approximation is given by
\bea
\frac{d^2\sigma}{d\Omega dE'} =  \frac{\alpha^2}{Q^4} \frac{E}{E'} L_{\mu \nu}W^{\mu \nu},
\eea
where $\Omega$ is the laboratory solid angle of the scattered electron, $E'$ is the energy of the scattered electron, $L_{\mu \nu}$ is the leptonic tensor
\bea
L_{\mu \nu} &=& 2 (k_\mu k'_\nu + k'_\mu k_\nu - g_{\mu \nu} k\cdot k').
\eea
In $L_{\mu\nu}$, $k^\mu$ and $k^{'\mu}$ denote the initial and final electron momenta respectively and $W^{\mu \nu}$ is the hadronic tensor given by
\bea
\label{Wmunu}
W^{\mu \nu } &=& \text{Im} \>T^{\mu \nu} , \qquad T^{\mu \nu } =  i \int d^4 x \> e^{i q\cdot x} \langle N| T\left(J_\mu(x)J_\nu(0)\right) |N\rangle ,
\eea  
where $J^\mu$ is the electromagnetic current of the struck quark and $q^\mu=k^\mu - k^{'\mu}$ with $q^2=-Q^2$. The above form of $W^{\mu \nu}$ follows from the optical theorem in which the imaginary part of the forward Compton amplitude is related to the cross-section for fully inclusive scattering off the initial state nucleon ($N$). Lorentz and gauge invariance dictate the following general form for the hadronic tensor
\begin{equation}\label{f3}
W_{\mu\nu}=\left(\frac{q^\mu q^\nu}{q^2}-g^{\mu\nu}\right)F_1(x,Q^2)+\left(P^{\mu}-\frac{P\cdot q}{q^2}q^\mu\right)\left(P^{\nu}-\frac{P\cdot q}{q^2}q^\nu\right)\frac{F_2(x,Q^2)}{\nu}, \nn \\
\end{equation}
where $P^\mu$ is the initial nucleon momentum, $\nu=P\cdot q$ and $F_{1,2}$ are dimensionless structure functions. The moments of these structure functions can be written in terms of the OPE as shown in Eq.(\ref{qh2}) for $F_2$.

The structure of the product of electromagnetic currents given in Eq.(\ref{Wmunu}) at light-like distances is given by Wilson's operator product expansion, see Refs. \cite{Wilson:1969zs, Zimmermann:1972tv, Okawa:1980ei} 
\bea\label{OPE}
J\left(\frac{x}{2}\right)J\left(-\frac{x}{2}\right)=\sum_{n=0}^{\infty}\sum_{i,\tau}\>C_{i,\tau}^{n}(x^2)\>\mathcal{O}^{i\tau,\>\mu_1\ldots\mu_n}_{n}(0)\>x_{\mu_1}\ldots x_{\mu_n}
\eea
where we have suppressed the Lorentz indices on the currents.  The operators appearing on the RHS of Eq.(\ref{OPE}) are symmetric and traceless in indices $\mu_1\ldots \mu_n$ and thus have a definite \textit{twist} (dimension - spin) denoted by $\tau$.  

The structure functions $F_i$ are determined by both the Wilson coefficients and the matrix elements of the operators in Eq.(\ref{OPE}), the Fourier transforms of $C_{i,\tau}^{n}(x^2)$ are related to the moments of the structure functions e.g.
\bea\label{moments}
\int_0^1\>\text{d}x\>x^{n-2}F_{1,2}(x,Q^2)\simeq\sum_jC_{j,\tau}^{n}(Q^2)\langle N| \mathcal{O}^{j,\tau}_n(0) | N\rangle .
\eea 

The dependence on $Q^2$ is controlled by the anomalous dimension of the operators in Eq.(\ref{moments}).  The bare (${\cal O}_{n}^{\tau i (b)}$) and renormalized (${\cal O}_{n}^{\tau i}$) operators of Eq.(\ref{qh2}) are related by 
\bea
{\cal O}_{n}^{\tau i(b)} = Z_{n\tau}^{ij} \>{\cal O}_{n}^{\tau j}, 
\eea
where $Z_{n\tau}^{ij}$ denote the renormalization constants.  They are, in general, matrices since different operators of a given spin $n$ mix under renormalization. From the scale invariance of the bare operators one can derive the RG evolution equations 
\bea
\mu \frac{d}{d\mu} {\cal O}_{n}^{\tau j} &=& - \gamma^{ji}_n \>{\cal O}_{n}^{\tau i} , \qquad \gamma^{ji} = Z_{n\tau}^{(-1)jk}\mu \frac{d}{d\mu}Z_{n\tau}^{ki},
\eea
and from the $\mu-$independence of the moments it follows that the Wilson coefficients satisfy the RG equation
\bea
\mu \frac{d}{d\mu} C_{n\tau}^{j} &=& \gamma^{ji}_n \>C_{n\tau}^{i},
\eea
which can be solved to give
%\bea
%\label{RG}
%C_{n\tau}^{i} (\mu, Q^2) &=& \text{P} \>\text{exp} \Big [ \int_{g(\mu_0)}^{g(\mu)}\frac{\gamma_n^\text{T}(g)}{\beta(g)} dg\Big ]_{ij}C_{n\tau}^{i} (\mu_0, Q^2),
%\eea
%where $\mu_0$ and $\mu$ denote the initial and final values for the RG running. Typically, $\mu_0^2 \sim Q^2$ and $\mu^2 \sim \Lambda^2_{QCD}$.
\bea
C_{n\tau}^{i}(Q^2/\mu^2,g)\simeq \sum_j\>C_{n\tau}^{j}(1,\bar{g}(t'))\>\text{T}\left[\text{exp}\left\{-\int_0^t\>\text{d}t'\>\gamma_{n\tau}(\bar{g}(t'))\right\}\right]_{ji} .
\eea
Where $t=1/2\>\text{ln}(Q^2/\mu^2)$, and $\bar{g}(t)$ is the running coupling in QCD.  In what follows, we will evaluate $\gamma_n^{ij}$ and its eigenvalues for the non-singlet, twist-4 operators.  Our main phenomenological task will then be an evaluation of Eq.(\ref{moments}) in the resonance region.

%%%%%%%%%%%%%%%%%%%%%%%%%%%%%%%%%%%%%%%
%												                            %
%				OPERATOR BASIS					                            %
%											                                     %
%%%%%%%%%%%%%%%%%%%%%%%%%%%%%%%%%%%%%%%
\section{Operator Basis}\label{basis}
In this section we review the basis of operators that appear at twist-2 and twist-4. At twist-2, it is well known that there are just two towers of operators for a given spin $n$.  The multiplicatively renormalizable flavor non-singlet (NS) operators are
\bea\label{eq:59}
\mathcal{O}^{\text{NS}}_{q;\mu_1\ldots\mu_n} &=& i^{n-1}\textbf{S}\left[\bar{\psi}_f\>\gamma_{\mu_1}D_{\mu_2}\ldots D_{\mu_n}\frac{\lambda_a}{2}\>\psi_f\right]-\text{trace terms},
\eea
and the two types of flavor-singlet operators that mix under renormalization are
\bea\label{eq:60}
\mathcal{O}^{\text{S}}_{q;\mu_1\ldots\mu_n} &=&i^{N-1}\textbf{S}\left[\bar{\psi}_f\>
\gamma_{\mu_1}D_{\mu_2}\ldots D_{\mu_n}\>\psi_f\right]-\text{trace terms}, \nn \\[1.2 em]
\mathcal{O}^{\text{S}}_{G;\mu_1\ldots\mu_n} &=& 2i^{N-2} \textbf{S}\left[F^a_{\mu_1\alpha}D_{\mu_2}\ldots D_{\mu_{n-1}}F^{\alpha,a}_{\mu_n}\right]-\text{trace terms}.
\eea
%\bea
%{\cal O}^{2f}_{\text{S},\mu_1 \cdots \mu_n} &=& \bar{\psi}_f \>\gamma_{\{\mu_1} (iD_{\mu_2})\cdots (iD_{\mu_n\}})\> \psi_f - \text{traces},\nn \\
%{\cal O}^{2g}_{\text{S},\mu_1 \cdots \mu_n} &=& -\frac{1}{2}F^{\{\mu_1 \alpha} (iD_{\mu_2}) \cdots (iD_{\mu_{n-1}}) F^{\mu_n \}}_\alpha - \text{traces},
%\eea
Here $\psi_f$ denotes a quark field of flavor $f$ and $F^{\alpha \beta}$ denotes the gluon field strength tensor and $\lambda_a$ is an SU(3)$_f$ genertor. The operation \textbf{S} reminds one to symmetrize the Lorentz indices in brackets.  The operators in Eqs.(\ref{eq:59}) and (\ref{eq:60}) are thus completely symmetric and traceless in the indices $\mu_1,\cdots \mu_n$ and transform under irreducible representations of the Lorentz group of spin-$n$. The anomalous dimension matrix for the flavor singlet operators takes the schematic form
\bea
 \gamma^{n} &=&   \left (  \begin{array}{cc}
a_{ff}^{n} & a_{fg}^{n}\\
a_{gf}^{n} & a_{gg}^{n} \end{array} \right ).
\eea
The diagonal entries $a_{ff}^{n}$ and $a_{gg}^{n}$ arise from self-renormalization graphs for operators ${\cal O}^S_{q;\mu_1 \cdots \mu_n}$
and ${\cal O}^S_{G;\mu_1 \cdots \mu_n}$ respectively. The off-diagonal entries come from graphs that mix these two operators.  The flavor non-singlet operator $\mathcal{O}^{\text{NS}}_q$ undergoes multiplicative renormalization since it cannot mix into the flavor-singlet operators ${\cal O}_q^{\text{S}},\>{\cal O}_G^{\text{S}}$.

The situation for twist-4 is more complicated. In general the operators at twist-4 can be classified into several types which mix at the one-loop level.  The specific number of operators grows with the spin-$n$ unlike the case at twist-2. A complete basis of twist-4 operators with tree level Wilson coefficients was first given in Refs.~\cite{Jaffe:1981td, Jaffe:1981sz, Jaffe:1982pm} and parts of the anomalous dimension matrix were computed at one loop in Refs.~\cite{Gottlieb:1978zj,Okawa:1980ei,Okawa:1981uw}. The `canonical' basis in Refs.~\cite{Jaffe:1981td, Jaffe:1981sz, Jaffe:1982pm} was constructed by requiring that the time-ordered product of electromagnetic currents is expanded in terms of operators that (a) are totally symmetric, (b) traceless, and (c) contain no contracted derivatives. In this paper, we extend the work of Refs.~\cite{Gottlieb:1978zj,Okawa:1980ei,Okawa:1981uw} and compute the full anomalous dimension matrix of flavor non-singlet operators at twist-4 and spin-2. The operators at twist-4 that contribute at spin-2 are given by

\bea
\label{jaffe-basis-1}
\Delta \cdot Q_n^{1(k,\ell)} &=& g \bar{\psi}_R\Deltaslash \>d^{^{\!\!\!\! \leftarrow \ell}}d^{^{\!\!\!\! \rightarrow k}} \psi _R \>\bar{\psi}_R \Deltaslash \> d^{^{\!\!\!\! \rightarrow n-2-k-\ell}}\psi _R, \nn \\
%%%%%
\Delta \cdot Q_n^{2(k,\ell)} &=& g \bar{\psi}_R \tau _a\Deltaslash \>d^{^{\!\!\!\! \leftarrow \ell}}d^{^{\!\!\!\! \rightarrow k}} \psi _R \>\bar{\psi}_R \Deltaslash \> d^{^{\!\!\!\! \rightarrow n-2-k-\ell}}\tau _a\psi _R, \nn \\
%%%%%%
\Delta \cdot Q_n^{3(k,\ell)} &=& g \bar{\psi}_R
\Deltaslash \>d^{^{\!\!\!\! \leftarrow \ell}}d^{^{\!\!\!\! \rightarrow k}} \psi _R \>\bar{\psi}_L \Deltaslash \> d^{^{\!\!\!\! \rightarrow n-2-k-\ell}}\psi _L, \nn \\
%%%%%%
\Delta \cdot Q_n^{4(k,\ell)} &=& g \bar{\psi}_R \tau _a\Deltaslash \>d^{^{\!\!\!\! \leftarrow \ell}}d^{^{\!\!\!\! \rightarrow k}} \psi _R \>\bar{\psi}_L \Deltaslash \> d^{^{\!\!\!\! \rightarrow n-2-k-\ell}}\tau _a\psi _L, \nn \\
%%%%%%
\Delta \cdot Q_n^{5(k,\ell)} &=& g \bar{\psi}_L \Deltaslash \>d^{^{\!\!\!\! \leftarrow \ell}}d^{^{\!\!\!\! \rightarrow k}} \psi _L \>\bar{\psi}_L \Deltaslash \> d^{^{\!\!\!\! \rightarrow n-2-k-\ell}}\psi _L, \nn \\
%%%%%%
\Delta \cdot Q_n^{6(k,\ell)} &=& g \bar{\psi}_L \tau _a \Deltaslash \>d^{^{\!\!\!\! \leftarrow \ell}}d^{^{\!\!\!\! \rightarrow k}} \psi _L \>\bar{\psi}_L \Deltaslash \> d^{^{\!\!\!\! \rightarrow n-2-k-\ell}}\tau _a \psi _L, \nn \\
%%%%%%
\Delta \cdot Q_n^{7(k)} &=&  \bar{\psi} \>d^{^{\!\!\!\! \leftarrow k}}\>\fslash ^{\!\!\!\!\! *} \>\gamma _5 d^{^{\!\!\!\! \rightarrow n-1-k}}\psi   , \nn \\
%%%%%%
\Delta \cdot Q_n^{8(k)} &=& i \bar{\psi} \Deltaslash \>d^{^{\!\!\!\! \leftarrow k}}\>\fslash  \> d^{^{\!\!
\!\! \rightarrow n-1-k}}\psi   , \nn \\
%%%%%%
\eea
where $\Delta$ is a light-like vector, $\Delta \cdot Q_n=\Delta^{\mu_1}\cdots \Delta^{\mu_n}Q_{n, \mu_1\cdots \mu_n}$, $d=i\Delta^\mu D_\mu$, $f^\beta = F^{\rho \beta}\Delta_\rho$, and $^{*}f^\beta=\epsilon^{\rho \beta \sigma \tau } F_{\sigma \tau}\Delta_\rho$. The subscripts $R,L$ on the quark fields denote the chirality so that $\psi_{R,L}= \frac{1\pm \gamma_5}{2}\psi$.   In this paper we compute the anomalous dimension matrix of the operators listed in Eq.(\ref{jaffe-basis-1}), however in the small $x$-Bjorken domain, we expect purely gluonic operators to make the main contribution.  For twist-4, in addition to the quark operators listed above, we list purely gluonic operators as well \cite{Bartels:1999km}.
\bea
\label{jaffe-basis-2}
\Delta \cdot O_n^{G1}&=& \text{Tr}[F^{\alpha \beta} \>d^{^{^{\!\!\!\! \rightarrow n}} } F_{\alpha \beta}] \nn \\
\Delta \cdot O_n^{G2(k,\ell)}&=& \text{Tr}[ f_\alpha d^{^{^{\!\!\!\! \rightarrow n-4-k-\ell}} } f^{\alpha} \> d^{^{^{\!\!\!\! \rightarrow k}} }f_{\beta}  d^{^{^{\!\!\!\! \rightarrow \ell}} }f^\beta] \nn \\
\Delta \cdot O_n^{G3 (k,\ell)}&=& \text{Tr}[ f_\alpha  d^{^{^{\!\!\!\! \rightarrow n-4-k-\ell}} }f^{\beta} d^{^{^{\!\!\!\! \rightarrow k}} } f_{\alpha}  d^{^{^{\!\!\!\! \rightarrow \ell}} }f^\beta] \nn \\
\Delta \cdot O_n^{G4 (k,\ell)}&=& \text{Tr}[ f_\alpha  d^{^{^{\!\!\!\! \rightarrow n-4-k-\ell}} }f^{\beta} d^{^{^{\!\!\!\! \rightarrow k}} } f_{\beta}  d^{^{^{\!\!\!\! \rightarrow \ell}} }f^\alpha] 
\nn \\
\Delta \cdot O_n^{G5 (\ell)}&=& \text{Tr}[ f_\alpha  d^{^{^{\!\!\!\! \rightarrow n-2-\ell}} } \>F_{\alpha\beta}\> d^{^{^{\!\!\!\! \rightarrow \ell}} } f^{\beta}]  
\eea

%%%%%%%%%%%%%%%%%%%%%%%%%%%%%%%%%%%%%%%
%												                            %
%				RENORMALIZATION 				                            %
%											                                     %
%%%%%%%%%%%%%%%%%%%%%%%%%%%%%%%%%%%%%%%
\section{Renormalization of Twist-4 Operators}
\label{anom-dim}

\begin{figure}
\begin{center}
\includegraphics[scale=1.2]{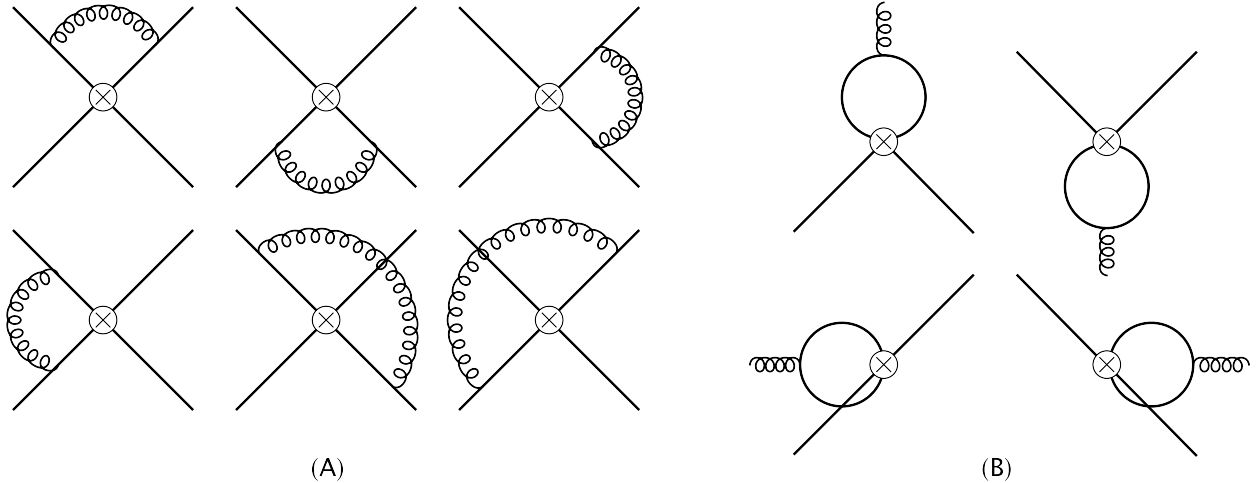}
\caption{Renormalization of 4Q operators.  The diagrams in (A) are self-renromalization graphs.  The graphs in (B) are annihilation graphs.}
\label{4qgraphs}
\end{center}
\end{figure}

\begin{figure}
\begin{center}
\includegraphics[scale=1.2]{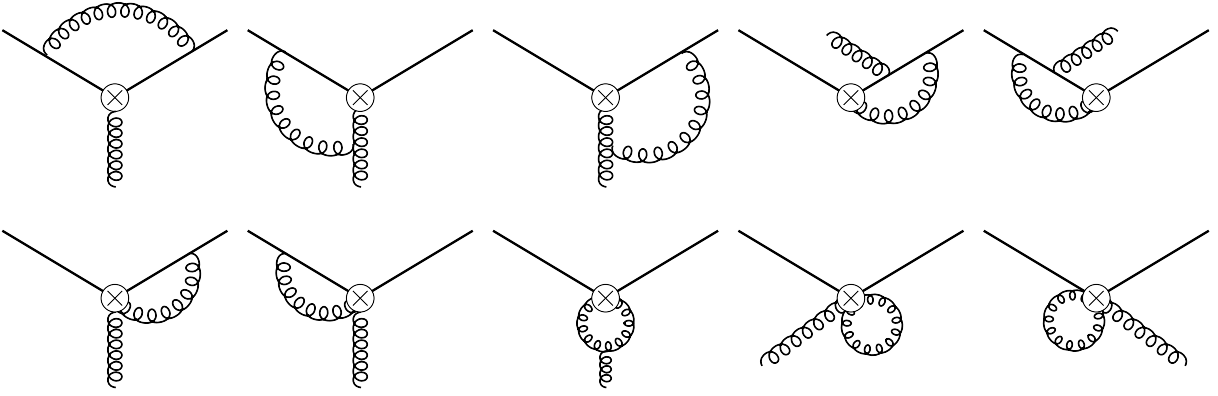}
\caption{Feynman diagrams for the renormalization of 2Q operators.}
\label{2qgraphs}
\end{center}
\end{figure}

\begin{figure}
\includegraphics{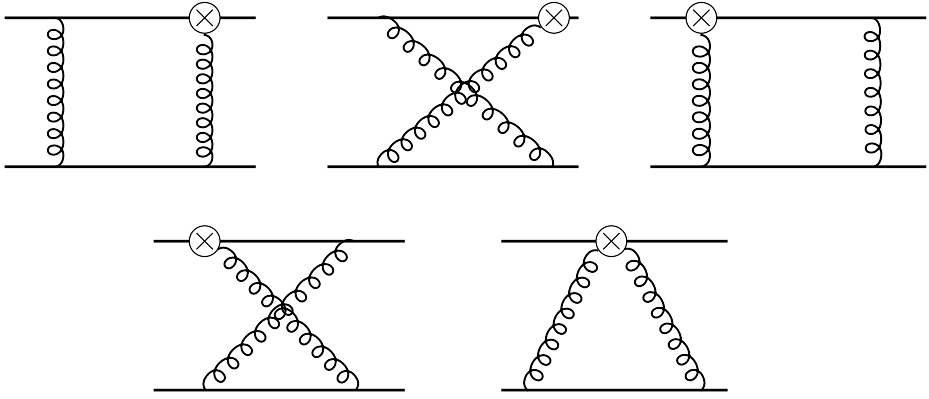}
\caption{Feynman diagrams for $2Q \rightarrow 4Q$ mixing.}
\label{boxgraphs}
\end{figure}

The renormalization of the twist-4 operators listed in the last section is complicated by the large number of mixings present.  In this section, we classify the operator basis into distinct types to better organize the calculation of the one-loop anomalous dimension matrix that determines the  RG evolution at leading order.  In Eqs.(\ref{jaffe-basis-1}) and (\ref{jaffe-basis-2}), the operators $Q^1_n - Q^6_n$ are 4-quark operators,  $Q_n^7, Q_n^8$ are 2-quark operators, and $Q_n^{G1} - Q_n^{G5} $ are pure gluon operators which we symbolically denote as 4Q, 2Q, and $G$ type operators respectively.  In terms of this classification, the anomalous dimension matrix then takes the following schematic form
%%%%%%%%%%%%%
\bea
\label{anom-schem}
\gamma_n &=&  \left ( \begin{array}{ccc}
\gamma^{4Q\to 4Q}_n\>\> & \gamma^{4Q\to 2Q}_n \>\>& \gamma^{4Q\to G}_n\\
\gamma^{2Q\to 4Q}_n \>\> & \gamma^{2Q\to 2Q}_n\>\> & \gamma^{2Q \to G}_n\\
\gamma^{G \to 4Q}_n \>\> &\gamma^{G\to 2Q}_n \> \> & \gamma^{G\to G}_n\end{array} \right ) ,
\eea
where $\gamma^{4Q\to 4Q}_n$ is a matrix that arises from the self-renormalization graphs of the 4Q operators, $\gamma^{2Q\to 4Q}_n$ denotes contributions from graphs that mix the 2Q  operators into 4Q operators and so on. Recall that we are restricting our analysis to spin-2 (n=2),  flavor non-singlet twist-4 operators. At one-loop, Fig.~\ref{4qgraphs}(A) shows the QCD self-renormalization graphs of the 4Q type operators.  The graphs of Fig.~\ref{4qgraphs}(B) contribute to the 4Q$\to$4Q self-renormalization after using the QCD equations of motion which were not considered in previous work.  Fig.~\ref{2qgraphs} shows the self-renormalization of the 2Q operators, and Fig.~\ref{boxgraphs} shows the 2Q$\to$4Q mixing graphs. For the graphs in Fig.~\ref{2qgraphs}, we have chosen to compute using the background field method~\cite{Abbott:1980hw}.  
%%%%%%%%%%%%

Below we review in schematic notation, the ingredients that go into the anomalous dimension calculation. The bare and renormalized operators are related as
\bea
\label{renorm-op-basis}
{\cal O}_b^i &=& Z^{ij} {\cal O}^j ,
\eea
where the subscript $b$ on the LHS indicates a bare operator and the operator on the RHS denotes the renormalized operator and $Z^{i j}$ denotes the renormalization constants.
The indices $i,j$ run over the basis of operators. We also denote the renormalization factors for the massless fermion wave function and the strong coupling constant as $Z_\psi$ and $Z_g$ respectively so that the bare ($b$) and renormalized quantities are related as

\bea
\psi_b = \sqrt{Z_\psi} \psi, \qquad g_b = \mu^\epsilon Z_g g.
\eea
The renormalization constants can be expanded around unity as
\bea
Z_\psi = 1+ \delta Z_\psi, \qquad Z_g &=& 1+ \delta Z_g,
\eea
where $\delta Z_\psi$ and $\delta Z_g$ denote the contributions from higher order perturbative diagrams.

We have found that the anomalous dimension for $\gamma^{4Q\to 2Q}$ is zero as one expects on general grounds~\cite{Braun:2009vc}. Consequently, only the $\gamma_2^{4Q\to 4Q}$, $\gamma_2^{2Q\to 2Q}$, and $\gamma_2^{2Q\to 4Q}$ blocks of the anomalous dimension matrix are relevant. We break the matrix $Z^{ij}$ of Eq.(\ref{renorm-op-basis}) into the component blocks $Z^{4Q,4Q},  Z^{2Q,2Q},$ and $Z^{2Q,4Q}$ corresponding to mixings among the 4Q operators, the 2Q operators, and the mixing of 2Q operators into 4Q operators respectively. As mentioned previously, the 4Q operators do not mix into the 2Q operators and since we restrict our analysis to flavor non-singlet operators we do not include the pure gluon G-type operators in the basis. The $Z^{4Q,4Q}$ and $ Z^{2Q,2Q}$ renormalization matrices can be expanded around the unit matrix as
\bea
Z^{2Q,2Q} &=& 1+ \delta Z^{2Q,2Q}, \qquad Z^{4Q,4Q} =1+ \delta Z^{4Q,4Q}, 
\eea
while the off-diagonal block $Z^{2Q,4Q}$ gets non-zero contributions starting at one-loop and is written as
\bea
Z^{2Q,4Q} &=& \delta Z^{2Q,4Q}.
\eea
We now have all the necessary notation to discuss the extraction of the one-loop anomalous dimension. We outline the steps for the 4Q
and 2Q operator renormalization below.

%----------------------------FOUR QUARK OPERATORS---------------------
\subsection{Four-Quark Operators}

The bare 4Q operators have the schematic form
\bea
{\cal O}_b^{4Q} = g_b^2 \> \bar{\psi}_b \psi_b \bar{\psi}_b \nn \psi_b,
\eea
where we have suppressed flavor indices and the Lorentz and Dirac structure.
The renormalized and bare 4Q operators are related as
\bea
{\cal O}^{4Q} &=& (Z^{-1})^{4Q, 4Q} \>{\cal O}_b^{4Q}  \nn \\
&=& g^2 \mu^{2\epsilon}\> \bar{\psi} \psi \bar{\psi}  \psi 
 + (2\delta Z_\psi + 2\delta Z_g - \delta Z^{4Q,4Q}) \> g^2 \mu^{2\epsilon}\> \bar{\psi}\psi \bar{\psi}  \psi,   \nn \\
 \eea
 where the second term above is just the counterterm and determines the renormalization matrix $\delta Z^{4Q,4Q}$. Once $\delta Z^{4Q,4Q}$ is extracted
 from the counterterm above, the anomalous dimension matrix is given by
 is given by
 \bea
\gamma^{4Q\to4Q} =  (Z^{4Q,4Q})^{-1} \mu \frac{d}{d\mu}Z^{4Q,4Q}.
\eea

%----------------------------TWO-QUARK OPERATORS---------------------
\subsection{2Q Operators}

The bare 2Q operators have the schematic form
\bea
{\cal O}^{2Q}_b = \bar{\psi}_b g_bF_b \psi_b ,
\eea
where $F_b$ denotes the bare field strength tensor and we have suppressed flavor indices and Lorentz and Dirac structure. There renormalized 2Q operator is related to the bare operators as
\bea
\label{2Qrenorm}
{\cal O}^{2Q} &=& (Z^{-1})^{2Q, 2Q} \>{\cal O}_b^{2Q} + (Z^{-1})^{2Q, 4Q} \>{\cal O}_b^{4Q} \nn \\
&=& \bar{\psi} g F \psi + (\delta Z_\psi - \delta Z^{2Q,2Q}) \>\bar{\psi} gF \psi +   (\delta Z^{-1})^{2Q,4Q} Z_\psi^2Z_g^2 \mu^{2\epsilon}  g^2 \> \bar{\psi} \Gamma \psi \bar{\psi} \Gamma \psi ,
\eea
where the two terms in the first line above correspond to mixing among the 2Q operators and the mixing of the 2Q operators into 4Q operators respectively. The combination $g_bF_b$ remains unrenormalized in the background field method, and the last two terms in the second line of Eq.(\ref{2Qrenorm}) denote the counterterms and the anomalous dimension components are given by
 \bea
\gamma^{2Q\to2Q} =  (Z^{2Q,2Q})^{-1} \mu \frac{d}{d\mu}Z^{2Q,2Q}, \qquad \gamma^{2Q\to4Q} =  (Z^{2Q,4Q})^{-1} \mu \frac{d}{d\mu}Z^{2Q,4Q}.
\eea
For one-loop renormalization, Eq.(\ref{2Qrenorm}) simplifies to
\bea
{\cal O}^{2Q} 
&=& \bar{\psi} g F \psi + (\delta Z_\psi^{(1)} - \delta Z^{(1)2Q,2Q}) \>\bar{\psi} gF \psi +   (\delta Z^{(1)2Q,4Q})^{-1} \mu^{2\epsilon}  g^2 \> \bar{\psi} \Gamma \psi \bar{\psi} \nn\Gamma \psi ,
\eea
where the superscript $(1)$ on the renormalization constants indicate the respective one-loop contributions.

%%%%%%%%%%%%%%%%%%%%%%%%%%%%%%%%%%%%%%%
%												                            %
%				POWER COUNTING					                            %
%											                                     %
%%%%%%%%%%%%%%%%%%%%%%%%%%%%%%%%%%%%%%%
\section{Consistent Power Counting in Powers of $g$}\label{counting}

\begin{figure}
\includegraphics{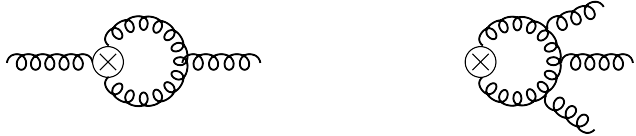}
\caption{ \textit{Left} - A representative Feynman diagram illustrating operator $G_4$ mixing into $G_3$,  an order $g$ correction to $G_3$.  \textit{Right} - The Feynman diagram illustrating the mixing of $G_3$ into $G_4$, which is order $g^3$. }
\label{CountingGraphs}
\end{figure}

Before proceeding to the calculation of the anomalous dimension, we address an  issue concerning consistent treatment of orders in perturbation theory \cite{Jaffe:1981td}.  Using abbreviated notation, we collectively call $O_1 = (\bar{\psi} \psi)^2$, $O_2 = \bar{\psi} F \psi$, $G_3 =F D^2 F$, and $G_4 = F^3 $.  A close look at the mixings among $O_1 - G_4$ reveals that $Z_{G_3 \rightarrow G_4}$ is order $g^3$ whereas $Z_{G_4 \rightarrow G_3}$ is order $g$ as shown in Fig.~\ref{CountingGraphs}.  One can readily see that the counting inconsistencies persist when computing the mixings $Z_{G_3\rightarrow O_1}$ as well.  It is desirable to write the anomalous dimension in a scaled form $\gamma_{ij}\simeq g^2 d_{ij}$ when computing Eq.(\ref{moments}).  This form of $\gamma$ renders direct calculations of the integral
\bea
\text{T}\> \text{exp}\left [- \int \frac{\tilde{\gamma}(g')}{\beta( g')} \right]
\eea 
to be quite simple.  However, a leading log evolution of the operators $O_1 - G_4$ leads to an anomalous dimension matrix which is not proportional to one consistent power in the coupling, e.g.
\bea
\tilde{\gamma}(g') \sim \left(
\begin{array}{cccc}
g^2 \>\gamma_{11} & g^3 \> \gamma_{12} & 0 & 0  \\
 g \>\gamma_{21} & g^2\> \gamma_{22} & g\> \gamma_{23}  & g^2\> \gamma_{24} \\
 0 & g^3\> \gamma_{32} & g^2\>\gamma_{33} & g^3\> \gamma_{34} \\
  0 & g^2\>\gamma_{42} & g\> \gamma_{43}  & g^2\> \gamma_{44}\\
\end{array}\right) .
\eea

A form of the mixing matrix proportional to $g^2$ in lowest order can be regained by an appropriate rescaling of the twist-4 operators.  We have chosen the following redefinitions
\begin{eqnarray*}
O_1 & \rightarrow &g^2 (\bar{\psi}\psi)^2,\\
O_2 &  \rightarrow & g (\bar{\psi} \slashed{F} \psi),\\
G_3 &  \rightarrow & F D^2 F,\\
G_4 &  \rightarrow & g F^3.
\end{eqnarray*}
Of course, the dominant logarithm is independent of such conventions.  After these redefinitions, the anomalous dimension matrix has a homogenous scaling in the strong coupling
\bea
\tilde{\gamma}(g') \sim g^2 \left(
\begin{array}{cccc}
\gamma_{11} &  \gamma_{12} & 0 & 0  \\
\gamma_{21} & \gamma_{22} & \gamma_{23}  &  \gamma_{24} \\
 0 &  \gamma_{32} & \gamma_{33} &  \gamma_{34} \\
  0 & \gamma_{42} &  \gamma_{43}  &  \gamma_{44}\\
\end{array}\right).
\eea
%In Figure 4 using the rescaled operators, for example, one can see that the left-most Feynman diagram is an order $g^2$ correction to $G_3$, and the rightmost graph is also an order $g^2$ correction to $G_4$ giving one overall power in the coupling.

%%%%%%%%%%%%%%%%%%%%%%%%%%%%%%%%%%%%%%%
%												                            %
%				FLAVOR STRUCTURE 				                            %
%											                                     %
%%%%%%%%%%%%%%%%%%%%%%%%%%%%%%%%%%%%%%%
\section{Flavor Structure}
\label{sec-flavor}
\begin{figure}
\includegraphics{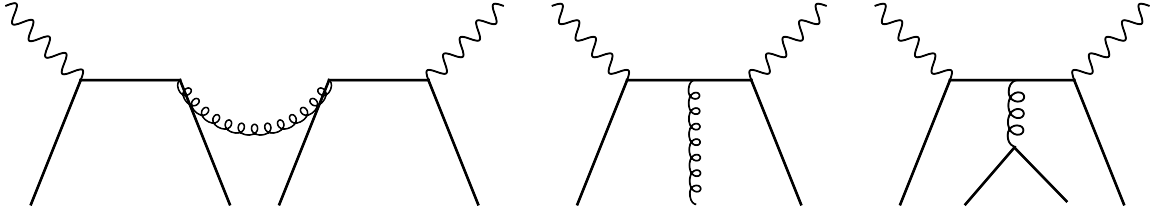}
\caption{\textit{Left} - Double hand bad diagram with flavor structure $\bar{\psi}Q\psi \bar{\psi}Q\psi$.\ \  \textit{Middle} - Feynman diagram with flavor structure $\bar{\psi}Q^2\psi$. \ \  \textit{Right} - Feynman diagram with flavor structure $\bar{\psi}Q^2\psi \bar{\psi}\psi$}
\label{twist-tree}
\end{figure}

In this section we discuss and establish notation for the flavor structure of the twist-4 operators. The structure of the anomalous dimension matrix can be organized according to  flavor structure since QCD with massless quarks preserves flavor symmetry. The electromagnetic current entering in the forward Compton amplitude 
in Eq.(\ref{Wmunu}) is given by

\begin{equation}\label{fs1}
J^\mu(x) = \bar{\psi}(x)\gamma^\mu Q \psi(x) , \qquad Q= \frac{1}{2}(\lambda^3 + \frac{1}{\sqrt{3}}\lambda^8),
\end{equation}
where $\psi$ is a column vector in flavor space so that $\psi = (\psi_u, \psi_d,\psi_s)$ and $Q$ is the electromagnetic
charge operator acting on $\psi$ and can be written in terms of the SU(3)$_f$ Gell Mann matrices $\lambda_i$  as shown.
The twist-4 operators from the OPE of the product of electromagnetic currents in Eq.(\ref{Wmunu}) can be classified in terms
of their transformation properties under SU(3)$_f$.  Schematically, the 4Q and 2Q operators have the following flavor structures
\bea
\label{flavor}
&&4Q:  A)\> \bar{\psi} Q \psi \bar{\psi} Q \psi, \qquad B) \>\bar{\psi} Q^2 \psi \bar{\psi} \psi ,\nn \\
&&2Q: C)\> \bar{\psi}Q^2\psi,
\eea
where the precise color and  Dirac  structure is suppressed. The flavor structure $A)$ in Eq.(\ref{flavor}) arises from  the first handbag diagram of Fig.~\ref{twist-tree}. The second diagram of Fig.~\ref{twist-tree} generates both $B)$ and $C)$ flavor structures where the flavor structure in $B)$ arises after an application the gluon equation of motion (EOM) for the external gluon.  These flavor structures can then be decomposed into irreducible representations of SU(3)$_f$ with definite isospin $(I,I_z)$ and hypercharge (Y=2$\lambda_8/\sqrt{3}$)\cite{deSwart:1963gc}. Since the charge operator $Q$ is a linear combination of $\lambda_3$ and $\lambda_8$ all these operators have $I_z=Y=0$.

The flavor decomposition of the 4Q operator of type A) in  Eq.(\ref{flavor}) is given by
\bea
\label{decomp1}
\bar{\psi} Q \psi \> \bar{\psi} Q \psi = \sqrt{\frac{2}{3}} O^{27,A}_{I=2} + \frac{2}{\sqrt{10}} O^{27,A}_{I=1} +  
\frac{2}{\sqrt{30}} O^{27,A}_{I=0}+
 \frac{2}{\sqrt{15}} O^{8,A}_{I=1} +  \frac{2}{3\sqrt{5}}
 O^{8,A}_{I=0}  -  \frac{\sqrt{2}}{3} O^{1,A}_{I=0} ,\nn \\
\eea
and the flavor structure $\bar{\psi} Q^2\psi$ that appears in the 4Q operators of type B) and 2Q operators of type C) in Eq.(\ref{flavor}) is decomposed as
\bea
\label{decomp2}
\bar{\psi} Q^2 \psi \bar{\psi}\psi &=& O^{8,B}_{I=1} + O^{8,B}_{I=0} \nn\\
\bar{\psi} Q^2 \psi &=&O^{8,C}_{I=1} + O^{8,C}_{I=0} .
\eea
The superscripts on the operators on the RHS of Eqs.(\ref{decomp1}) and (\ref{decomp2}) 
 denote the SU(3)$_f$ representation and the subscripts
denote the isospin representation of the SU(2)$_I$ subalgebra of
SU(3)$_f$.  The labels $(A,B,C)$ are included to remind the reader the specific flavor structure given by Eq.(\ref{flavor}).  For notational convenience we define ``meson" fields that make the flavor structure of quark bilinears manifest as
\bea
\label{pions}
\pi ^+ &=& \bar{d}u \>\>\>\>\>\>\>\>\>\> \>\>\>\>\>  \>\>\>\>\> \>\>\>\>\> \>\>\>\>\> \>K^+ = \bar{s}u \>\>\>\>\>\>\>\>\>\>  \bar{K}^0 = \bar{d}s
\>\>\>\>\>\>\>\>\>\> \eta _8 = -\frac{1}{\sqrt{6}} (\bar{d}d+\bar{u}u -2\bar{s}{s})\nn \\
\pi ^0 &=& \frac{1}{\sqrt{2}} (\bar{d}d-\bar{u}u) \>\>\>\>\>\>\>\>\>\>  K^0 = \bar{sd}
\>\>\>\>\>\>\>\>\>\> K^- = -\bar{u}s  
\>\>\>\>\>\>\>\> \eta' = \frac{1}{\sqrt{3}} (\bar{u}u+\bar{d}d +\bar{s}{s})\nn \\
\pi^- &=& -\bar{u}d. \nn 
\eea
In terms of these fields, the flavor structure of the operators of the type A operators appearing on the RHS of Eq.(\ref{decomp1}) is given by
\bea
\label{flavor-ops}
O^{27,A}_{I=2} &\sim & \frac{1}{\sqrt{6}} [ 2\pi ^0 \pi ^0 + \pi ^+ \pi ^- + \pi ^-
\pi ^+ ] \nn \\
O^{27,A}_{I=1} &\sim&   \frac{1}{\sqrt{10}} [ (\bar{K}^0K^0 +  K^-K^+) +
(K^0\bar{K}^0 + K^+ K^-)  + \sqrt{3} (\pi ^0 \eta _8 + \eta
_8 \pi ^0 ) ]\nn \\
O^{27,A}_{I=0} &\sim&  \frac{3}{\sqrt{30}} \> [ \>\frac{1}{6} \>(\pi ^0 \pi
^0 - \pi ^+\pi ^- -\pi ^- \pi ^+ ) - \frac{1}{2} (\bar{K}^0 K^0 - K^-K^+ )
  -\frac{1}{2} (K^0\bar{K}^0 - K^+K^- )
 + \frac{3}{2} \eta _8 \eta _8]
\nn \\ 
O^{8,A}_{I=1} &\sim& -\sqrt{\frac{3}{5}} \> [ \> \frac{1}{2} (\bar{K}^0K^0 + K^-K^+
 ) + \frac{1}{2} (K^0\bar{K}^0 +K^+ K^-  ) -\frac{1}{\sqrt{3}} (\pi ^0 \eta _8 + \eta
_8 \pi ^0 )] \nn
\\
O^{8,A}_{I=0} &\sim&  \frac{1}{\sqrt{5}} [(\pi ^0 \pi
^0 - \pi ^+\pi ^- - \pi ^- \pi ^+  ) -\frac{1}{2} (\bar{K}^0 K^0 - K^-K^+)
  -\frac{1}{2} (K^0\bar{K}^0 - K^+K^-  )
 - \eta _8 \eta _8]  \nn \\ 
O^{1,A}_{I=0} &\sim&  -\frac{1}{\sqrt{8}} [ ( \pi ^0 \pi ^0 - \pi ^+\pi ^- -\pi ^- \pi ^+  )   
 + (\bar{K}^0K^0  - K^-K^+ )+(K^0\bar{K}^0 - K^+K^-) 
 + \eta _8 \eta _8 ]. \nn\\ 
\eea
The last operator $O^{1,A}_{I=0}$ in Eq.(\ref{flavor-ops}) is a flavor singlet and can mix with the pure gluon operators and will not be considered
in the rest of the analysis. The flavor structure of operators of type B in Eq.(\ref{decomp2}) are given by
 \bea\label{flavor-ops2}
O^{8,B}_{I=1} &\sim& -\frac{1}{\sqrt{6}}\pi^0\eta'\nn
\\
O^{8,B}_{I=0} &\sim&  -\frac{1}{\sqrt{18}} \eta_8\eta' +\frac{2}{3}\eta'\eta',
\eea
and the flavor structure of the two quark operator $\bar{\psi}Q^2\psi$ is,
\bea\label{flavor-ops3}
O^{8,C}_{I=1} &\sim& -\frac{2}{3\sqrt{2}}\> \pi^0 \nn\\
O^{8,C}_{I=0} &\sim& \frac{2}{3\sqrt{3}} \eta' -\frac{1}{3\sqrt{6}}\eta_8.
\eea 

%%%%%%%%%%%%%%%%%%%%%%%%%%%%%%%%%%%%%%%
%												                            %
%				ANOMALOUS DIMENSION			                       %
%											                                     %
%%%%%%%%%%%%%%%%%%%%%%%%%%%%%%%%%%%%%%%
\section{Structure of the Anomalous Dimension Matrix}
\label{anom-dim-flavor}

In this section, we expand the discussion of section ~\ref{anom-dim} on the structure of the anomalous dimension matrix to incorporate the flavor structure discussed in section \ref{sec-flavor}. 
Equations (\ref{decomp1}-\ref{flavor-ops3}) give the SU(3) flavor decomposition of the 4Q and 2Q type operators. The conservation of flavor in massless QCD implies that the $O^{27,A}_{I=2,1,0}$ operators in Eq.(\ref{decomp1}) will not mix with  operators living in a different representation of SU(3) or with those in a different isospin subgroup. On the other hand, the octet operators $O^{8,A}_{I=1}$ and $O^{8,B,C}_{I=1}$  can mix with each other.  %\footnote{For economy of notation will often replace the symbol $8\to 8$ when referring to operators of type A.}. 
Thus, the analog of Eq.(\ref{renorm-op-basis}) that relates the bare and renormalized operators for $O^{27}_{I=2,1,0}$ (dropping the A label) is diagonal
\bea
\left(
\begin{array}{c}
 \vec{O}^{27}_{I=2} \\
 \vec{O}^{27}_{I=1} \\
\vec{O}^{27}_{I=0} \\

\end{array}
\right)_b=\left(
\begin{array}{cccc}
\mathbb{P}_{I=2} & 0 & 0   \\
 0 & \mathbb{P}_{I=1} & 0   \\
 0 & 0 & \mathbb{P}_{I=0} \\
\end{array}
\right)\left(
\begin{array}{c}
 \vec{O}^{27}_{I=2} \\
 \vec{O}^{27}_{I=1} \\
 \vec{O}^{27}_{I=0} \\
 \end{array}
\right),
\eea
where the vector $ \vec{O}^{27}_{I}$  is a 6-dimensional column vector corresponding to the Dirac and color structures of the 4Q operators $Q_{n=2}^{1,\cdots, 6}$ of Eq.(\ref{jaffe-basis-1}) with flavor structure given by  the \textbf{27} flavor representation with isospin $I$ appearing in Eq.(\ref{decomp1})
\bea
 \vec{O}^{27}_{I} &=& (Q_{n=2}^{1(0,0)}, Q_{n=2}^{2(0,0)}, Q_{n=2}^{3(0,0)}, Q_{n=2}^{4(0,0)}, Q_{n=2}^{5(0,0)}, Q_{n=2}^{6(0,0)})^T_{27,I},
\eea
 where the superscript $T$ denotes the transpose.
 Note that there are no 2Q operators in the \textbf{27} representation of SU(3)$_f$. The renormalization constants $\mathbb{P}_{I} $ are thus $6\times 6$ matrices. The anomalous dimension matrix for the \textbf{27} operators is given by
\bea
\gamma_{I}^{27} &=& \mathbb{P}_{I}^{-1} \mu \frac{d}{d\mu}  \mathbb{P}_{I}.
\eea

The octet sector of the anomalous dimension matrix is more complicated.  Operator types $(A,B,C)$ in Eq.(\ref{flavor}) all contain flavor-octet operators, and in general, these three operator types will mix under renormalization.  For flavor structures of type $A)$ and $B)$ one encounters only four quark operators while for $C)$ one has the two-quark operators. For convenience, we embed those of type $A)$ in a ten-component vector $\vec{O}^{8,A}_{I}$ and combine those of type $B)$ and $C)$ into a second ten component vector,  $\vec{O}^{8,BC}$.  The first 6 entries of $\vec{O}^{8,BC}$ are filled by the $Q^{8,B}$ (4Q) operators and the last four operators are filled by $Q^{8,C}$ (2Q) operators,
\bea
\vec{O}^{8,A}_{I} &=&  (Q_{n=2}^{1(0,0)}, Q_{n=2}^{2(0,0)}, Q_{n=2}^{3(0,0)}, Q_{n=2}^{4(0,0)}, Q_{n=2}^{5(0,0)}, Q_{n=2}^{6(0,0)}, 0, 0, 0, 0)^T_{8,A,I}, \medskip\nn \\
\vec{O}^{8,BC}_{I} &=&  (Q_{n=2}^{1(0,0)}, Q_{n=2}^{2(0,0)}, Q_{n=2}^{3(0,0)}, Q_{n=2}^{4(0,0)}, Q_{n=2}^{5(0,0)}, Q_{n=2}^{6(0,0)}, Q_{n=2}^{7(0)},Q_{n=2}^{7(1)},Q_{n=2}^{8(0)},Q_{n=2}^{8(1)})^T_{8,BC,I}.\nn\\
\eea
With these definitions, the bare and renormalized operators are related as
\bea
\left(
\begin{array}{c}
 \vec{O}^{8,A}_{I} \medskip \\
 \vec{O}^{8,BC}_{I} 
 \end{array}
\right)_b=\left(
\begin{array}{cc}
  \mathbb{L}_{I} & \mathbb{M}_{I} \medskip\\
 \mathbb{Q}_{I} & \mathbb{N}_{I} \\
 \end{array}
\right)\left(
\begin{array}{c}
 \vec{O}^{8,A}_{I} \medskip\\
 \vec{O}^{8,BC}_{I} \\
 \end{array}
\right).
\eea
%In this case, the vector $\vec{O}^{8,A}_{I}$ is a 6-dimensional vector composed of 4Q operators, and the vector $\vec{O}^{8,BC}$ is a 10-dimensional vector including the six 4Q operator structures arising at tree-level from flavor structure $B)$ and the four 2Q operator structures at spin-2 arising from flavor structure $C)$. For convenience, we embed the six operators 
%as seen in Eq.(\ref{jaffe-basis-1})
%We have included the zeros in $\vec{O}^{8,A}$ for convenience so that it has the same dimension as $\vec{O}^{8,B,C}$.  

The matrices $\mathbb{L,M,N,Q}$ are then $10 \times 10$ matrices which have the form 
\bea
\mathbb{L}_{I}&=&\left(
\begin{array}{cc}
\mathbb{L}_{I}^{4Q\rightarrow 4Q} &\mathbb{L}_{I}^{4Q\rightarrow 2Q}  \\
\mathbb{L}_{I}^{2Q\rightarrow 4Q} & \mathbb{L}_{I}^{2Q\rightarrow 2Q}  \\
\end{array}
\right)
\qquad
\mathbb{N}_{I=1}=
\left(
\begin{array}{cc}
\mathbb{N}_{I}^{4Q\rightarrow 4Q} & \mathbb{N}_{I}^{4Q\rightarrow 2Q}  \\
\mathbb{N}_{I}^{2Q\rightarrow 4Q} & \mathbb{N}_{I}^{2Q\rightarrow 2Q} \\
\end{array}
\right)\nn \\
%%%%%%
\mathbb{M}_{I}&=&\left(
\begin{array}{cc}
\mathbb{M}_{I}^{4Q\rightarrow 4Q} & \mathbb{M}_{I}^{4Q\rightarrow 2Q}  \\
\mathbb{M}_{I}^{2Q\rightarrow 4Q}  & \mathbb{M}_{I}^{2Q\rightarrow 2Q} \\
\end{array}
\right)
\qquad
\mathbb{Q}_{I}=\left(
\begin{array}{cc}
\mathbb{Q}_{I}^{4Q\rightarrow 4Q} & \mathbb{Q}_{I}^{4Q\rightarrow 2Q}  \\
\mathbb{Q}_{I}^{2Q\rightarrow 4Q}  & \mathbb{Q}_{I}^{2Q\rightarrow 2Q} \\
\end{array}
\right)
\eea
The $\mathbb{L}_{I}$ and $\mathbb{N}_{I}$ matrices encode the remormalization structure of the 4Q and 2Q operator structures for the $8AB$ and $8C$ representations respectively.
The matrices $\mathbb{M}_{I}$ and $\mathbb{Q}_{I}$ encode the mixing of the 4Q and 2Q operator structures between the $8AB$ and $8C$ representations. Many of the submatrices
in $\mathbb{L}_{I},\mathbb{M}_{I},\mathbb{N}_{I},\mathbb{Q}_{I}$ vanish
\bea
\mathbb{L}_{I}^{4Q\rightarrow 2Q} &=& \mathbb{L}_{I}^{2Q\rightarrow 4Q} =\mathbb{L}_{I}^{2Q\rightarrow 2Q} = \mathbb{M}_{I}^{4Q\rightarrow 2Q}= \mathbb{N}_{I}^{4Q\rightarrow 2Q}= \mathbb{Q}_{I}^{4Q\rightarrow 2Q}= 0
\eea
All but the $\mathbb{L}_{I}^{4Q\rightarrow 2Q}$ vanish since only $4Q$ operators appear in $\vec{O}^{8,A}_{I}$. The remaining sub-blocks  $\mathbb{M}_{I}^{4Q\rightarrow 2Q}$, $\mathbb{N}_{I}^{4Q\rightarrow 2Q}$, $\mathbb{Q}_{I}^{4Q\rightarrow 2Q}$ would give rise of mixing of $4Q$ into $2Q$ operators (see Section~\ref{apxA} for more details). However, it is known on general grounds that such mixing does not arise (see {\em e.g.}, Ref.~\cite{Braun:2009vc}), a result that we have reproduced with our explicit computation. 

%%%%%%%%%%%%%%%%%%%%%%%%%%%%%%%%%%%%%%%
%												                            %
%				WILSON COEFFICIENTS				                            %
%											                                     %
%%%%%%%%%%%%%%%%%%%%%%%%%%%%%%%%%%%%%%%
\section{Wilson Coefficients for $n=2$}\label{wcs}
Having outlined the flavor structures of the twist-4 contributions, we now present the Wilson Coefficients for the twist-4 2Q and 4Q operators at leading spin. According to the formalism established in Ref.\cite{Jaffe:1981td}, the Compton amplitude at twist-4 naturally divides into two pieces arising from the graphs of Fig.~\ref{twist-tree}
\bea
\label{eq:correlator1}
-i \int d^4 x \> e^{i q\cdot x} \> \text{T} \left[ J_\mu(x) J_\nu(0)\right] &=& X_{\mu
\nu}+Y_{\mu\nu} .
\eea
The $Y_{\mu \nu}$ term arises from the double handbag-type diagrams (corresponding to the first and last diagram in Fig.~\ref{twist-tree}) and the $X_{\mu \nu}$ term arises from the remaining diagrams.
The explicit calculations of $X_{\mu\nu}$ and $Y_{\mu\nu}$ are given in detail in Ref.\cite{Jaffe:1981td}, and we summarize the full form of these expressions in appendix B.  For a leading moment ($n=2$) analysis, these expressions simplify greatly.  %For the leading moment $n=2$ and at tree-level, 
The $Y_{\mu \nu}$ term is given by
\bea\label{yeq}
Y_{\mu\nu}^{T=4,n=2} &=& - \frac{4g}{q^6}\>T^{\mu_1\mu_2}_{\mu\nu} \>\mathcal{O}^{2(0,0)}_{n=2,\mu_1\mu_2}\nn\\
T_{\mu\nu}^{\mu_1\mu_2} &=& q^2 g^{\mu_1}_\mu g^{\mu_2}_{\nu} -(g^{\mu_1}_\mu q_\nu+g^{\mu_1}_\nu q_\mu)q^{\mu_2} +g_{\mu\nu}q^{\mu_1}q^{\mu_2},
\eea
and the $X_{\mu \nu}$ term is given by
\bea\label{xeq}
X_{\mu\nu}^{T=4,n=2} &=& -\frac{g}{2 q^6} \left [ \frac{q^\mu q^\nu}{q^2}-g^{\mu\nu}\right ]\left\{2\>q\cdot\mathcal{O}^{7(0)}_{n=2}-3\> q\cdot\mathcal{O}^{3(0)}_{n=2}-3\>q\cdot\mathcal{O}^{3(1)}_{n=2}\right\} \nn \\
&-&\frac{g}{2 q^6}\left[  g^{\mu\nu} - \frac{p^\mu q^\nu+p^\nu q^\mu}{p \cdot q} + \frac{q^2 p^\mu p^\nu}{(p\cdot q)^2} \right]\left\{\frac{1}{2}\>q\cdot\mathcal{O}^{3(0)}_{n=2}+\frac{1}{2}\>q\cdot\mathcal{O}^{3(1)} + 5\>q\cdot\mathcal{O}^{7(0)}_{n=2}\right\}. \nn \\
\eea
Here $q\cdot \mathcal{O}$ is shorthand for $q_{\mu_1}\ldots q_{\mu_n} \mathcal{O}^{\mu_1\ldots \mu_n}$, while the explicit form of the operators appearing in $X_{\mu\nu}$ and $Y_{\mu\nu}$ in terms of the canonical operators of Eq.(\ref{jaffe-basis-1}) is\cite{Jaffe:1981td} 
\bea\label{treeops}
\Delta\cdot \mathcal{O}^{2(0,0)}_{n=2} &=& \Delta^\textbf{.} Q^{2(0,0)}_{n=2}-2\>\Delta ^\textbf{.}Q^{4(0,0)}_{n=2}+\Delta^\textbf{.} Q_{n=2}^{6(0,0)} \nn \\
\Delta\cdot\mathcal{O}^{3(0)}_{n=2} &=& \> \> \> \> \Delta^\textbf{.}Q^{7(0)}_{n=2} \nn \\
\Delta\cdot\mathcal{O}^{3(1)}_{n=2} &=& - \> \Delta^\textbf{.}Q^{7(1)}_{n=2} \nn \\
\Delta\cdot\mathcal{O}^{7(0)}_{n=2} &=& \Delta^\textbf{.} Q^{2(0,0)}_{n=2}+2\>\Delta ^\textbf{.}Q^{4(0,0)}_{n=2}+\Delta^\textbf{.} Q_{n=2}^{6(0,0)} .
\eea
Note that the first term in the RHS of Eq. (\ref{xeq}) contributes to the longitudinal structure function $F_L$ while all other terms in Eqs. (\ref{yeq},\ref{xeq}) contribute to $F_2$ whose moments we analyze in this work.

The foregoing decomposition does not yet reflect any flavor structure. To avoid introducing unwieldy notation, we simply indicate this structure below:
\bea
\label{eq:opflav}
\Delta\cdot \mathcal{O}^{2(0,0)}_{n=2}&&\qquad\text{Flavor Structure A: \textbf{27},$\textbf{8}$} \nn\\ 
\Delta\cdot\mathcal{O}^{7(0)}_{n=2} && \qquad\text{Flavor Structure B: $\textbf{8}$} \nn\\
\Delta\cdot\mathcal{O}^{3(0,1)}_{n=2}&& \qquad\text{Flavor Structure C: $\textbf{8}$}\ \ \ .
\eea

%The 4Q operators $\mathcal{O}^{2(0,0)}$ and  $\mathcal{O}^{7(0)}$ above transform under different flavor representations.  Specifically, the operator $\mathcal{O}^{2(0,0)}$ transforms in the \textbf{27} and $\textbf{8}_1$ of flavor as a 4Q operator of type $A$ ($\bar{\psi}Q \psi \bar{\psi}Q \psi$) shown in Eq.(\ref{decomp1}), while the operator $\mathcal{O}^{7(0)}$ transforms in the \textbf{8} of flavor as an operator of type $B$ ($\bar{\psi}Q^2 \psi \bar{\psi}\psi$). 

We now apply these results to the forward matrix elements of the vector current correlator in Eq.(\ref{eq:correlator1}). Several steps are required: (i) expressing the matrix elements in terms of those of the individual twist-four operators  $Q^{i,(k,l)}_{\mu_1\mu_2}$; (ii) decomposing the current-current product in terms of the various flavor structures $A)$, $B)$, and $C)$; and (iii) expressing the latter in terms of the operators associated with their SU(3) flavor decomposition given in Eqs.~(\ref{decomp1},\ref{decomp2}).  Starting with the first of these steps, we write the matrix elements of the twist-4 canonical basis operators as
\bea\label{me}
\langle \text{N} |\>Q^{i,(k,l)}_{\mu_1\mu_2}\> | \text{N} \rangle= \mathcal{A}^{i,(k,l)}\left( p_{\mu_1} p_{\mu_2} - \frac{1}{4}M_N^2 g_{\mu_1\mu_2}\right)
\eea
where the $\mathcal{A}$ factors are reduced matrix elements encoding the non-perturbative multi-parton correlations.  Second, using this form for the matrix elements in Eqs.~(\ref{yeq})  and (\ref{xeq}), we write the $n=2$ component of $T_{\mu\nu}$ as 
%\bea\label{fca}
%T_{\mu\nu}&=&-i \int d^4 x \> e^{i q\cdot x} \> \langle P | \>\text{T} \left( J_\mu(x) J_\nu(0)\right) \>|P\rangle \nn\\
%&=&-\frac{\omega^2\>d_{\mu\nu}}{Q^2}\left\{\underbrace{(\mathcal{A}^{2,(0,0)}-2\>\mathcal{A}^{4,(0,0)}+\mathcal{A}^{6,(0,0)})}_{\bar{\psi}Q\psi\bar{\psi}Q\psi}+ \frac{5}{8}\>\underbrace{(\mathcal{A}^{2,(0,0)}+2\>\mathcal{A}^{4,(0,0)}+\mathcal{A}^{6,(0,0)})}_{\bar{\psi}Q^2\psi\bar{\psi}\psi} \right. \nn\\ 
%&&\qquad\qquad+\left. \frac{1}{16}\>\underbrace{(\mathcal{A}^{7,(0)}-\mathcal{A}^{7,(1)})}_{\bar{\psi}Q^2\psi}\right\}\nn\\
%&-&\frac{\omega^2\>e_{\mu\nu}}{Q^2}\left\{\frac{1}{4}\underbrace{(\mathcal{A}^{2,(0,0)}+2\>\mathcal{A}^{4,(0,0)}+\mathcal{A}^{6,(0,0)})}_{\bar{\psi}Q^2\psi\bar{\psi}\psi}-\frac{3}{8}\underbrace{(\mathcal{A}^{7,(0)}-\mathcal{A}^{7,(1)})}_{\bar{\psi}Q^2\psi}\right\},
%\eea
\bea\label{fca}
T_{\mu\nu}&=&-i \int d^4 x \> e^{i q\cdot x} \> \langle P | \>\text{T} \left[ J_\mu(x) J_\nu(0)\right] \>|P\rangle\Bigr\vert_{n=2} \nn\\
\nn\\
&=&-\frac{\omega^2\>d_{\mu\nu}}{Q^2}\left\{\mathcal{A}_A+ \frac{5}{8}\>\mathcal{A_B} + \frac{1}{16}\>\mathcal{A}_C\right\}-\frac{\omega^2\>e_{\mu\nu}}{Q^2}\left\{\frac{1}{4}\mathcal{A}_B-\frac{3}{8}\mathcal{A}_C\right\},
\eea
in agreement with Ref.\cite{Choi:1993cu}. Here,  we have defined $\omega^2 = 1/x_B^2$ and  have indicated the flavor structures of each matrix element for clarity, introducing the shorthand notation:
\bea\label{shorthand}
\mathcal{A}_A&\equiv& \mathcal{A}^{2,(0,0)}-2\>\mathcal{A}^{4,(0,0)}+\mathcal{A}^{6,(0,0)}\qquad\text{Flavor Structure: } \>\bar{\psi}Q\psi\bar{\psi}Q\psi\nn\\
\mathcal{A}_B&\equiv& \mathcal{A}^{2,(0,0)}+2\>\mathcal{A}^{4,(0,0)}+\mathcal{A}^{6,(0,0)}\qquad\text{Flavor Structure: } \>\bar{\psi}Q^2\psi\bar{\psi}\psi\nn\\
\mathcal{A}_C&\equiv& \mathcal{A}^{7,(0)}-\>\mathcal{A}^{7,(1)}\qquad\qquad\qquad\>\>\>\>\>\>\>\>\text{Flavor Structure: }\>\bar{\psi}Q^2\psi.
\eea
The tensors $e_{\mu\nu}$ and $d_{\mu\nu}$ are written in full in section~\ref{apxB}.  The coefficients of $d_{\mu\nu}$ contribute only to the $F_2$ structure function whereas the coefficients of $e_{\mu\nu}$ contribute to the $F_L$ structure function \cite{Jaffe:1981td}.  

We now express the leading moment of the isovector part of $F_2$ in terms of the foregoing matrix elements. In doing so, we also carry out the SU(3) decomposition following \cite{Gottlieb:1978zj}. The result is
\bea\label{moment}
M^{I=1}_{n=2,\tau=4}\left(Q^2\right) &=& \int\text{d}x_{B} \>F_{2,\tau=4}^{I=1}(x_B,Q^2) =\>\> \frac{1}{2Q^2}\sum_j\Bigl\{ \frac{2}{\sqrt{10}}C_A^{27,j}(Q^2)\mathcal{A}_{A,j}^{27}\nn\\
&+& \frac{2}{\sqrt{15}}C_A^{8,j}(Q^2)\mathcal{A}_{A,j}^{8}
+ C_B^{8,j}(Q^2)\>\mathcal{A}_{B,j}^{8}+ C^{8,j}_C(Q^2)\> \mathcal{A}_{C,j}^{8}\Bigr\}\\
\nn\\
&\equiv& M_{27}(Q^2) + M_{8A}(Q^2) + M_{8B}(Q^2)+M_{8C}(Q^2).
 \eea   
We have introduced the notation $\mathcal{A}^{27}$ for the matrix element of $O^{27}_{I=1}$ appearing in Eq.(\ref{flavor-ops}).  The subscripts A, B, C indicate the specific flavor structure of each operator, e.g. that $\mathcal{A}_B$ is the matrix element of an operator of type $\bar{\psi}Q^2\psi\bar{\psi}\psi$ and $\mathcal{A}_C$ is the matrix element of type $\bar{\psi}Q^2\psi$ respectively\footnote{For the flavor {\bf 27}, the subscript $A$ is clearly redundant, as only the structure $A$ can yield a {\bf 27} after SU(3$)_f$ decomposition. However, we retain the subscript in this case for overall uniformity of notation.}. The explicit factors of $2/\sqrt{10}$ result from the SU(3) decomposition of the $A,B,C$ type operators while the $C_A^{27,j}(Q^2)$ are the corresponding Wilson coefficients. The index $j$ runs over all relevant operators in the canonical basis in Eq.(\ref{jaffe-basis-1}) \cite{Jaffe:1981td}. The values of the Wilson coefficients at an appropriate input scale (discussed below) are given in Table~\ref{tab:wilson}. 

\begin{table}
\caption[]{Tree-level Wilson Coefficients $C_k^{N,j} (Q_0^2 )$ evaluated at the input scale $Q_0$. Here, $N$ denotes the SU(3) multiplet while $j$ runs over the set of isovector canonical operators in Eq.(\ref{jaffe-basis-1}).  We have not included Wilson Coefficients for operators with $j=8(0)$ and $j=8(1)$ as these coefficients all vanish at the input scale.)
\label{tab:wilson}}
\begin{center}
\begin{ruledtabular}
\begin{tabular}{ccccccccr}
$C_k^{N,j}$	& $j=1$ &$j=2$  &$j=3$ &$j=4$ &$j=5$ &$j=6$&$j=7(0)$ &$ j=7(1)$  \\
\hline
\\
$C_A^{27,j}  $   &  0  &  1  &  0  &  $-2$  &  0  &  1   & 0 & 0\\
$C_A^{8,j}  $   &  0  &  1  &  0  &  $-2$  &  0  &  1   & 0 & 0  \\
$C_B^{8,j}  $   &  0 &  5/8  &  0  & $ 5/4$  &  0  &  5/8   & 0 & 0 \\
$C_C^{8,j}  $  &0 &0 &0 &0 &0 &0 &  1/16  &  $-1/16$      \\
\end{tabular}
\end{ruledtabular}
\end{center}
\end{table}

 One may ask whether it is possible to isolate experiementally the 27-plet and octet  contributions to the leading moment.  To this end, we note that for unpolarized DIS processes, the photon couples to a current we denote by $\mathcal{J}_{\text{EM}} = V_3 +\sqrt{\frac{1}{3}}V_8$, where $V$ signifies the vector nature of the current, and we have defined $V_i^\mu= \bar{\psi} \frac{\lambda_i}{2}\gamma^\mu \psi$.   Following \cite{Gottlieb:1978zj}, a similar flavor decomposition can be done for the electroweak charged current.  The charged current contains both a strangeness-changing piece and a non-strangeness-changing piece.  For the strangeness-changing part of the charged current, the isovector 27-plet and octet moments are expressed as linear combinations of moments of $F_2$ extracted from neutral and charged current DIS processes (see Ref.\cite{Gottlieb:1978zj} for details) so that we can write   
\bea\label{flavormoment1}
M^{27}_{I=1,n=2}= \sqrt{\frac{1}{10}}\left\{3 (M^{\text{ep}}_{2}-M^{\text{en}}_{2}) - \frac{1}{2 \>\text{sin}^2\theta_\text{C}}(M_{2}^{\nu \text{p}}+M_{2}^{\bar{\nu}\text{p}}-M_{2}^{\nu\text{n}}-M_2^{\bar{\nu}\text{n}})\right\} ,
\eea
and
\bea\label{flavormoment2}
M^{8}_{I=1,n=2}=\frac{\sqrt{15}}{10}\left\{2 (M^{\text{ep}}_{2}-M^{\text{en}}_{2})  + \frac{1}{2 \>\text{sin}^2\theta_\text{C}}(M_{2}^{\nu \text{p}}+M_{2}^{\bar{\nu}\text{p}}-M_{2}^{\nu\text{n}}-M_2^{\bar{\nu}\text{n}})\right\}. 
\eea
%The Nachtmann moment $M_n(Q^2)$ of structure function $F_2$ is given by
%\bea
%M_n(Q^2) = \int_0^1 \> dx_{\text{B}} \> \frac{\xi^{n+1}}{x_{\text{B}}^3} F_2(x_{\text{B}},Q^2) \> \frac{3+3(n+1)r+n(n+2)r^2}{(n+2)(n+3)}
%\eea
%where $\xi = \frac{2x_{\text{B}}}{(1+r)}$ is the Nachtmann variable, and $r=\sqrt{1+\frac{4M^2x_{\text{B}}^2}{Q^2}}$ with $M$ being the mass of the nucleon.  Note that as $Q^2 \rightarrow \infty$, we regain the usual $x_{\text{B}}$ moment
%\bea
%\int_0^1 \> dx_{\text{B}} \> x_{\text{B}}^{n-2} F_2(x_{\text{B}},Q^2).
%\eea
%\bea
 %\sqrt{\frac{1}{10}}\left\{3 (F_{2,\gamma}^{\text{ep}}-F_{2,\gamma}^{\text{en}}) - \frac{1}{2 \text{sin}^2\theta_\text{C}}(F_{2,\text{CC}}^{\text{ep}}-F_{2,\text{CC}}^{\text{en}})\right\} =\mathcal{O}^{27}_{I=1} 
%\eea
%\bea
 %\frac{\sqrt{15}}{10}\left\{2 (F_{2,\gamma}^{\text{ep}}-F_{2,\gamma}^{\text{en}}) + \frac{1}{2 \text{sin}^2\theta_\text{C}}(F_{2,\text{CC}}^{\text{ep}}-F_{2,\text{CC}}^{\text{en}})\right\} =\mathcal{O}^{8}_{I=1} 
%\eea
Thus,  through a combination of experiments it is in principle possible to isolate specific flavor structures that contribute at twist-4 to the moments of $F_2$.

%%%%%%%%%%%%%%%%%%%%%%%%%%%%%%%%%%%%%%%
%												                            %
%				RENORMALIZATION GROUP 			                            %
%											                                     %
%%%%%%%%%%%%%%%%%%%%%%%%%%%%%%%%%%%%%%%
\section{Leading Log RG Evolution}\label{renorm}
%With both the perturbative and non-perturbative input, we now have the necessary ingredients to compute the isovector, flavor \textbf{27} and \textbf{8}  contributions to the moment of $F_2$ in the resonance region.  Rewriting Eq.(\ref{moments}) we have
%\bea\label{fm}
%M_{n=2}^{\tau=4}|_{I=1}(Q^2) = \>  \sum_{i,f=\textbf{27,8}} \frac{C^{(i)}_{n=2,f,I=1}(Q^2/\mu^2,g)}{Q^2} \> \mathcal{A}_{i}^{f}
%\eea
%where $\mathcal{A}_i^{f}$ includes the proper Clebsch-Gordon coefficients written down in Eqs.(\ref{flavor}, \ref{decomp1}) for the flavor \textbf{27} and \textbf{8} representations.  
%Rewriting Eq.(\ref{moments}) to include the $Q^2$ dependence of the moment, 
%\bea\label{fm}
%M_{n=2}^{\tau=4}|_{I=1}(Q^2) = \>  \sum_{i,f=\textbf{27,8}} \frac{C^{(i)}_{n=2,f,I=1}(Q^2/\mu^2,g)}{Q^2} \> \mathcal{A}_{i}^{f}
%\eea
%where $\mathcal{A}_i^{f}$ includes the proper Clebsch-Gordon coefficients written down in Eqs.(\ref{flavor}, \ref{decomp1}) for the flavor \textbf{27} and \textbf{8} representations.  
In this section we present results for the Wilson coefficients of the flavor non-singlet twist-4 operator combinations. Within each flavor representation, the RG evolution is affected by mixing between various operators in the canonical basis. The evolution of the Wilson coefficients in flavor representation $R$ is then given by
\bea
\label{evo-1}
C_i^R\left(\frac{Q^2}{\mu^2},g(t)\right) \> = \> \sum_j \> C_j^R(1,g(0)) \> \text{Exp}\Big[- \int_0^t dt' \>\gamma [g(t')] \Big]_{ji},
\eea
where $t=1/2 \ln (Q^2/\mu^2)$ and the subscripts $i,j$ label the Wilson coefficients of the canonical operators in Eq.(\ref{jaffe-basis-1}). The Wilson coefficients $ C_j^R(1,g(0))$ correspond to the values obtained in the matching calculation at $\mu^2=Q^2$. For leading log running, the $C_j^R(1,g(0))$ correspond to the tree level values obtained from the OPE. The Wilson coefficient on the LHS of  Eq.(\ref{evo-1}) corresponds to the value of the Wilson coefficient after RG evolution from the initial scale $Q^2$ to the final scale $\mu^2$. With one loop running the evolution of the strong coupling is given by
\bea
g^2(t) = \frac{g^2(0)}{1+ 2 \beta_0 g^2(0) t},
\eea
where $\beta_0\equiv1/(4\pi)^2(11/3 \>C_A-4/3 T_f\>n_f)$ and $g(0)$ corresponds to the strong coupling evaluated at $\mu^2=Q^2$. To solve the evolution equation, we first diagonalize the one-loop anomalous dimension matrix so that
\bea
\gamma_{ji}[g(t)] = g^2(t)  \> R_{jm} \>d_{m\ell} \>R^{-1}_{\ell i},
\eea
where $d_{m \ell} = \delta_{m\ell} d_m$ is the diagonalized matrix with eigenvalues
$d_m$ and $R$ denotes the appropriate rotation matrix.  The anomalous dimension matrix elements $\gamma_{ji}$ for the various canonical operators in different flavor representations are presented in section \ref{anom-dim-flavor}. The evolution equation for the Wilson coefficients can now be written as
\bea
C_i^R\left(\frac{Q^2}{\mu^2},g(t)\right) \> = \> \sum_{j,m} \> C_j^R(1,g(0)) \> \text{Exp}\Big[- \int_0^t dt' \>\frac{g^2(0) d_m}{1+ 2 \beta_0 g^2(0) t'}\Big] R_{jm} R^{-1}_{mi},
\eea
The Wilson coefficients are evolved from the scale $Q^2$ to the scale of the non-perturbative matrix elements which we denote as $\mu^2=Q^2_0$ and refer to as the input scale.  Thus, in terms of the latter the $Q^2$ dependence of the Wilson coefficients is given by
\bea
\label{evo-final}
C_i^R\left(\frac{Q^2}{Q_0^2},g(t_0)\right) \> = \> \sum_{j,m} \> C_j^R(1,g(0)) \> \text{Exp}\Big[- \int_0^{t_0} dt' \>\frac{g^2(0) d_m}{1+ 2 \beta_0 g^2(0) t'}\Big] R_{jm} R^{-1}_{mi},
\eea
where  $t_0=1/2 \ln (Q^2/Q_0^2)$ and $C_j^R(1,g(0))$ are the tree-level values determined from the OPE at the matching scale $Q_0^2=Q^2$.

The above expression can be simplified further to give
\bea
C_i^R\left(\frac{Q^2}{Q_0^2},g(t_0)\right) \> = \> \sum_{j,m} \> C_j^R(1,g(0)) \> \Big [ 1 + \beta_0 g^2(0) \ln \frac{Q^2}{Q_0^2} \Big ]^{-\frac{d_m}{2\beta_0}}R_{jm} R^{-1}_{mi},
\eea
where 
\bea
g^2(0) = \frac{1}{\beta_0 \ln \frac{Q^2}{\Lambda^2}}
\eea
so that 
\bea\label{coef}
C_i^R\left(\frac{Q^2}{Q_0^2},g(t_0)\right) \> = \> \sum_{j,m} \> C_j^R(1,g(0)) \> \Bigg [ 1 +  \frac{\ln (Q^2/Q_0^2)}{\ln (Q^2/\Lambda^2)} \Bigg ]^{-\frac{d_m}{2\beta_0}}R_{jm} R^{-1}_{mi},
\eea
%WE NEED TO PLOT THIS EQUATION AS A FUNCTION OF $\mu^2$ TO SHOW THE EFFECT OF RG RUNNING IN A GIVEN EXPERIMENT CONDUCTED AT FIXED $Q^2$. IN THE NEXT SECTION, WE WILL SET $\mu=Q_0$ TO SEE THE FINAL EFFECT OF RG RUNNING FROM $Q^2$ to $Q_0^2$ AND ITS IMPACT ON PHENOMENOLOGY>

\begin{figure}
\begin{center}
\includegraphics[scale=0.5]{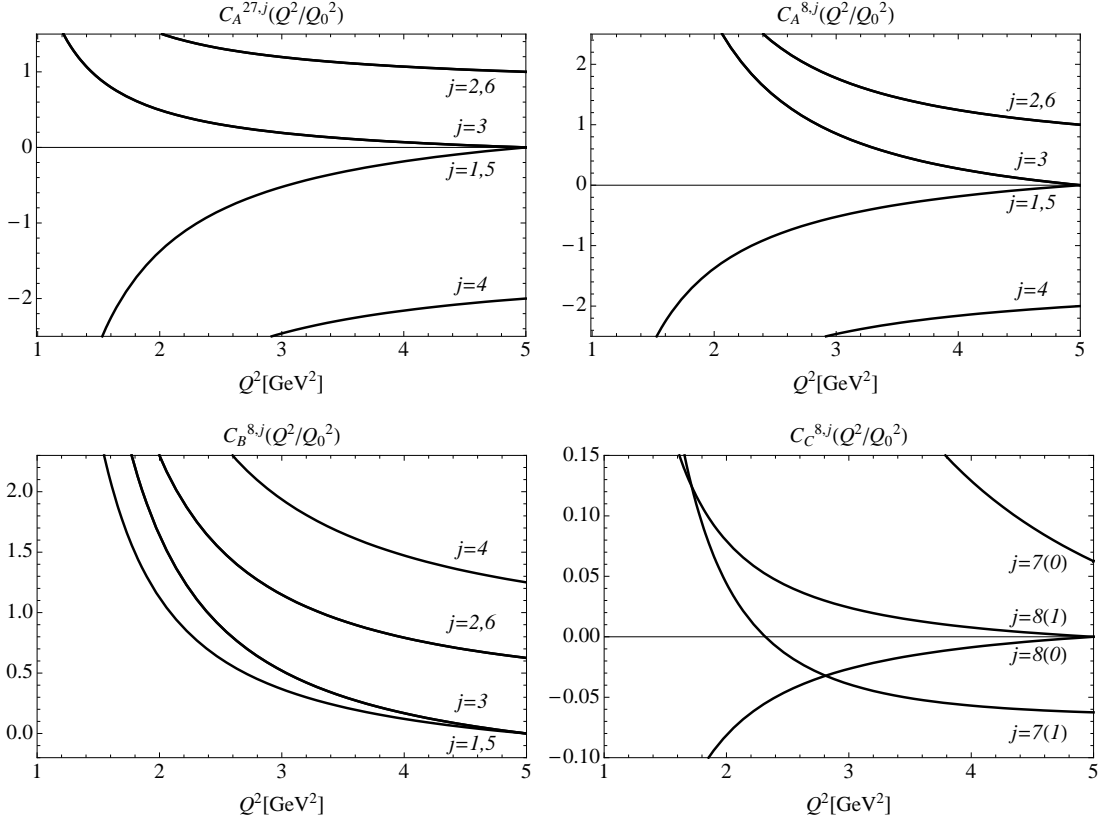}
\caption{Evolution of Wilson coefficients using Eq.(\ref{coef}).  The flavor representation of each coefficient is indicated at the top of each plot (see Eq.(\ref{moment})).  As input to Eq.(\ref{coef}), we have taken $\Lambda = 0.3\>\text{GeV}$ and $n_f=3$. }%We note the opposite sign running for $C_A^{27}$ and $C_A^8$.  The notation $C_{2,6}$ illustrates that both coefficients $C_2$ and $C_6$ have the same $Q^2$ dependence.}
\label{WCs}
\end{center}
\end{figure}

%The isovector flavor \textbf{27} and \textbf{8} contributions to the leading moment of $F_2$ are given by the equation 
%\bea \label{octetmoment}
%\int\text{d}x_{B} \>F_{2,\tau=4}^{I=1}(x_B,Q^2) &=&\frac{1}{2Q^2}\sum_j\left\{C^{j,27}_A(Q^2/\Lambda^2,g)\mathcal{A}_{A,j}+C^{j,8_1}_A(Q^2/\Lambda^2,g)\mathcal{A}_{A,j}^{8_1} \right.\nn\\
%&&\qquad\qquad\left.+C_B^j(Q^2/\Lambda^2,g)\mathcal{A}_{B,j}^{8_1}+C^j_C(Q^2/\Lambda^2,g)\mathcal{A}_{C,j}^{8_1}\right\},\nn\\
%\eea
%where each $C^j_{A,B,C}$ has the form given by Eq.(\ref{coef}), and the labels $(A,B,C)$ denote various flavor structures of the twist-4 matrix elements.  
In Fig.~\ref{WCs}, we plot the RG evolution of the Wilson coefficients in the isovector combination. We have used the input scale $Q^2_0= 5$ GeV$^2$, corresponding to the scale at which existing theoretical evaluations of the non-perturbative matrix elements have been performed (see Section \ref{matrixelements} below) and the tree-level Wilson coefficients given in Table~\ref{tab:wilson}.  We have chosen $n_f=3$ in Fig.~\ref{WCs} which is consistent with the SU(3)$_f$ decomposition. %and have checked that dropping heavy quark effects in the evolution yields negligible changes in the evolution. 
At the top of each plot, the notation $C_{A,B,C}^R$ denotes the Wilson coefficient in representation $R$ and of the type $A,B,C$ corresponding to flavor structures $\bar{\psi}Q\psi\bar{\psi}Q\psi$,  $\bar{\psi}Q^2\psi \bar{\psi}\psi$, and $\bar{\psi}Q^2\psi$ respectively.  Note that we evolve the coefficients to smaller rather than larger values of $Q^2$ since we are interested in the kinematic regime relevant to the CLAS analysis and since the scale $Q_0^2$ associated with existing theoretical matrix element input lies above this region. 

Several features emerge from Fig.~\ref{WCs}. First, several of the Wilson coefficients that vanish at the input scale become non-vanishing and significant in magnitude at lower scales due to operator mixing. Second, the coefficient $C_C^{8,j}$ for $j=7(1)$ changes sign, going through zero near $Q^2=2$ GeV$^2$. All other coefficients increase in magnitude with decreasing $Q^2$. We find no satisfying explanation for the particular behavior of $C_C^{8,j}$ for $j=7(1)$ other than that its tree-level value is rather small and that there exist various contributions from operator mixing having opposite signs. Together, these features lead to the possibility that significant cancellations among various operator contributions may occur at one scale -- leading to a suppressed twist-four effect -- but that these cancellations are broken at other scales by the $Q^2$ evolution. Alternately, the overall twist-four contribution to the leading moment may be relatively small at all scales due to suppressed values of the operator matrix elements at the input scale. We explore these possibilities in the following sections. 

\section{Higher Twist Matrix Elements - Theoretical Input}\label{matrixelements}

In order to obtain a prediction for $M^{I=1}_{n=2,\tau=4}\left(Q^2\right)$, we now require values of the hadronic matrix elements of the twist-four operators at the input scale $Q_0$ (see Eq.(\ref{moment})).
Knowledge of higher twist quark and gluon correlators in hadrons is of fundamental interest in order to understand the structure of baryons and mesons on the basis of QCD.  These matrix elements are between hadronic states, making model independent computations challenging.  However, there do exist attempts in the literature to compute twist-4 matrix elements at leading spin on the lattice \cite{Gockeler:2001xw}.  Due to the complicated mixings with lower dimensional operators however, the analysis performed in \cite{Gockeler:2001xw} was restricted to operators of the specific flavor channels outlined in section \ref{sec-flavor}. An alternate phenomenological approach was used in Ref.~\cite{Choi:1993cu}. In what follows, we draw on the results from these two studies to determine non-perturbative input for an initial, illustrative analysis of $M^{I=1}_{n=2,\tau=4}\left(Q^2\right)$.

Before proceeding we observe that the aforementioned matrix element calculations provide only partial input for the evaluation of the RHS of Eq.(\ref{moment}). The lattice study of Ref. ~\cite{Gockeler:2001xw} gives only a value for the combination $\mathcal{A}_A^{27}$ appearing in Eq.(\ref{shorthand}), at the input scale, while the work of Ref.~\cite{Choi:1993cu} gives only the octet contributions $\mathcal{A}_{B,C}^{8}$. Away from the input scale, one no longer has the specific linear combinations given in Eq.(\ref{shorthand}), owing to the evolution of the Wilson coefficients. A robust prediction for the evolution of $M^{I=1}_{n=2,\tau=4}\left(Q^2\right)$ would require values for the individual contributions to $\mathcal{A}_{A,B,C}$. Moreover, we could find no evaluation of the octet contribution $\mathcal{A}_A^{8}$, nor do there appear to exist any computations of the matrix elements whose Wilson coefficients vanish at the input scale but become non-zero at lower scales. Consequently, we will adopt some reasonable {\em ansatz} for the individual matrix elements, motivated by the computations that do exist but with the caveat that a complete computation will require values for all matrix elements. 

To proceed, we first consider the combination $\mathcal{A}_A^{27}$. The 4Q operators introduced in Ref.~\cite{Gockeler:2001xw} have been evaluated on the lattice using Wilson fermions in the quenched approximation.  Phenomenological constraints for twist-4 2Q operators were estimated in Ref.~\cite{Choi:1993cu}.  %Following the notation outlined in  \cite{Gockeler:2001xw}, the twist-4 contribution in the $F_2$ structure function for the \textbf{27} representation is denoted 
%\bea
%A_{\mu\nu}^{27,\textbf{I=1}} =\bar{\psi}\gamma_\mu\gamma_5\tau^a\psi\bar{\psi}\gamma_\nu\gamma_5\tau^a\psi.
%\eea
%Expanding this operator in terms of canonical operators appearing in Eq.(\ref{jaffe-basis-1}) 
%\bea
%A_{\mu\nu}^{27,\textbf{I=1}} = Q^{2,(0,0)}_{\mu\nu}-2Q^{4,(0,0)}_{\mu\nu}+Q^{6,(0,0)}_{\mu\nu},
%\eea
%and taking the reduced matrix element yields the linear combination for the $\textbf{27}$ as shown in Eq.(\ref{fca}).  
The reduced matrix element of the 4Q operator $\mathcal{A}_A^{27}$ shown in Eq.(\ref{moment}) at an input scale of $Q_0^2\approx5 \> \text{GeV}^2$ was computed on the lattice and found to be 
\bea\label{4qme}
\mathcal{A}^{27,\textbf{I=1}}_A\Bigr\vert_{\text{Latt}}=(-10.4\pm 1.6)\times 10^{-4}\>\text{GeV}^2. 
\eea
We take this to be the value of the reduced matrix element $\mathcal{A}^{27}_A$ appearing in Eq.(\ref{moment}).  Neglecting the $Q^2$-dependence associated with evolution, the corresponding contribution to the Nachtmann moment for $F_2$ at leading spin is then \cite{Gockeler:2001xw} 
\bea
M^{27,\,I=1}_{n=2,\tau=4}\left(Q^2\right)=\int_0^1\text{d}x\>F_{2,\text{Latt}}^{\textbf{27},I=1}(x,Q^2) = -0.001(1)\frac{m_p^2\alpha_s}{Q^2}.
\eea
We note that the matrix element of $A_{\mu\nu}^{27,\textbf{I=1}}$ causes the moment to be quite small relative to the flavor non-singlet twist-2 operator, whose corresponding contribution is 0.14 at the same input scale (see Refs.\cite{Detmold:2002ri, Dolgov:2002zm}).

We could find no corresponding lattice computation of $\mathcal{A}^{8}_A$ which appears in Eq.(\ref{moment}).  As a benchmark, we first take its value to be equal to $\mathcal{A}^{27}_A$ at the input scale $Q_0^2\approx 5 \>\text{GeV}^2$, though  a larger magnitude for $\mathcal{A}^{8}_A$ is possible (see below). This leaves the twist-4, 4Q operator of type $B)$
%$\bar{\psi}Q^2\psi \bar{\psi}\psi$ 
and the 2Q operator of type $C)$. 
%$\bar{\psi}Q^2\psi$,
We were also unable to find lattice calculations of these operator matrix elements in the literature. As an alternative, we use the phenomenological estimates  obtained in Ref.\cite{Choi:1993cu} that rely, in part,  on information extracted from DIS data from both CERN and SLAC.  The 2Q and 4Q operators studied in Ref.\cite{Choi:1993cu} have the forms
\bea
Q^{g}_{\mu\nu}&=&ig\>\bar{\psi}\left\{D_\mu,\tilde{F}_{\nu\alpha}\right\}\gamma^\alpha\gamma_5 Q^2\psi,\\
\nn\\
Q^2_{\mu\nu}&=&g^2\bar{\psi}\gamma_\mu Q^2\tau^a \psi\>\bar{\psi}\gamma_\nu \tau^a\psi.
\eea
After integrating by parts, the 2Q operator $Q^g$  is the tree-level 2Q combination for the octet combination appearing in Eq.(\ref{fca}): 
\bea\label{q7rel}
Q^g_{\mu\nu} &=& ig\>\bar{\psi}\left(D_\mu\tilde{F}_{\nu\alpha}+\tilde{F}_{\nu\alpha}D_\mu\right)\gamma^\alpha\gamma_5Q^2\psi\nn\\
&\rightarrow&ig\>\bar{\psi}\left(-\overleftarrow{D}_\mu\tilde{F}_{\nu\alpha}+\tilde{F}_{\nu\alpha}\overrightarrow{D}_\mu\right)\gamma^\alpha\gamma_5 Q^2\psi\nn\\
&=&Q^{7,(0)}_{\mu\nu}-Q^{7,(1)}_{\mu\nu}.
\eea  
Following \cite{Choi:1993cu}, we denote the reduced matrix elements these operators $\mathcal{A}^g$ and $\mathcal{A}^2$, respectively.  For the proton (neutron) they are written,
\bea
\mathcal{A}^g_{p(n)}&=&Q_u^2\>K_{u(d)}^g+Q_d^2\>K_{d(u)}^g,\\
\mathcal{A}^2_{p(n)}&=&Q_u^2\>K_{u(d)}^2+Q_d^2\>K_{d(u)}^2,
\eea
where
\bea
K_{u(d)}^{g}&=&\frac{2ig}{M^2}\langle P| \bar{u}\left\{D_+,\tilde{F}_{+\mu}\right\} \gamma^\mu\gamma^5 \>u|P\rangle, \\
K^2_{u}&=& \frac{2}{M^2}\langle P | (\bar{u} \gamma_+ \tau^a u)\>(\bar{u}\gamma_+\tau^a u) | P \rangle,\\
K^2_d&=&\frac{2}{M^2}\langle P | (\bar{d} \gamma_+ \tau^a d)\>(\bar{d}\gamma_+\tau^ad) | P \rangle,
\eea
and $\gamma_+=1/\sqrt{2}(\gamma_0+\gamma_3)$.  The neutron matrix elements can be obtained from isospin symmetry and the strangeness contribution has been neglected to simplify the analysis.  The range of possible values for $K_u^g$ and $K_u^2$ (see Ref.~\cite{Choi:1993cu} for details) are
\bea
-0.585\>\text{GeV}^2 \le\> &K_{u}^g& \> \le -0.238\>\text{GeV}^2\\
-0.318\>\text{GeV}^2 \le \> & K_u^2 & \> \le 0.203\>\text{GeV}^2.
\eea
The values for $K_{d}^{g,2}$ were then computed by introducing a flavor ansatz, e.g.
\bea
K_{d}^{g,2}/K_{u}^{g,2}\simeq\frac{\int\>\text{d}x(d(x)+\bar{d}(x))x}{\int\>\text{d}x(u(x)+\bar{u}(x))x}\equiv\beta.
\eea
Where $u(x),d(x)$ are the twist-2 parton distribution functions and $\beta=0.476$ at $Q^2=5\>\text{GeV}^2$.  Finally, the isovector combination corresponds to the difference $\mathcal{A}_{p}^{g,2}-\mathcal{A}_{n}^{g,2}$ the final results are,
\bea
\nn
\mathcal{A}^{8}_{B}&\equiv& \mathcal{A}_{p}^{2}-\mathcal{A}_{n}^{2}= \frac{1}{3}(1-\beta)K_{u}^{2} \\
\label{eq:hatsuda}
\mathcal{A}^{8}_{C}&\equiv& \mathcal{A}_{p}^{g}-\mathcal{A}_{n}^{g}= \frac{1}{3}(1-\beta)K_{u}^{g}
\eea
from the range of $K_{u}^{g,2}$, the final values for the reduced matrix elements appearing in Eq.(\ref{moment}) are
\bea\label{mebc}
-0.10\>\text{GeV}^2 \le &\mathcal{A}^{8}_C& \le -0.04\>\text{GeV}^2\nn\\
-0.06\>\text{GeV}^2 \le &\mathcal{A}^{8}_B&\le  0.04 \>\text{GeV}^2.
\eea
For $\mathcal{A}^{8}_C$, we take the central value in this range.  However for $\mathcal{A}^{8}_B$, it is possible to choose a range of values both positive and negative and have chosen three representative cases, $\mathcal{A}_B = -0.06\>\text{GeV}^2$, $\mathcal{A}_B = 0.0\>\text{GeV}^2$ and $\mathcal{A}_B=0.04\>\text{GeV}^2$.   We assume a theoretical uncertainty in each value to be of the order of the lattice uncertainty for the 4Q matrix element $\mathcal{A}^{27}_A$ in Eq.(\ref{4qme}). A summary of all matrix element inputs is given in Table~\ref{tab:mes}.

\begin{table}
\caption[]{Reduced matrix elements used in computing $M^{I=1}_{n=2,\tau=4}\left(Q^2\right)$. Note that the input for the $\mathcal{A}_A^{8} $ represent two different {\em ansatz}. \label{tab:mes}}
\begin{center}
\begin{ruledtabular}
\begin{tabular}{llcr}
Operator	& Flavor Rep.  &Value  & Method \\
\hline
\\
$\mathcal{A}_A^{27}  $   & $\psi Q \psi \psi Q \psi $   & $-10.4\times10^{-4}\>\mathrm{GeV}^2$                                 & Lattice~\cite{Gockeler:2001xw}          \\
\\
$\mathcal{A}_A^{8}    $   & $\psi Q \psi \psi Q \psi $   & $-10.4\times 10^{-4}\>\mathrm{GeV}^2$                                &   $\mathcal{A}_A^{8} \sim \mathcal{A}_A^{27}
\vert_\mathrm{Latt}$ \\
\\
$\mathcal{A}_A^{8}    $   & $\psi Q \psi \psi Q \psi $   & $-10.4\times 10^{-3}\>\mathrm{GeV}^2$                                &   $\mathcal{A}_A^{8} \sim 10\times \mathcal{A}_A^{27}\vert_\mathrm{Latt}$ \\
\\
$\mathcal{A}_B^8       $   & $\psi Q^2\psi \psi \psi $   &  $-0.06 \le \mathcal{A}_B \le 0.04 \> \mathrm{GeV}^2$  & Phenomenology~\cite{Choi:1993cu} \\
\\
$\mathcal{A}_C^8       $   & $\psi Q^2 \psi$                  & $-0.1 \le \mathcal{A}_C \le -0.04\>\mathrm{GeV}^2$     & Phenomenology~\cite{Choi:1993cu} \\
\end{tabular}
\label{tme}
\end{ruledtabular}
\end{center}
\end{table}

We now observe that in order to predict the logarithmic corrections to the $1/Q^2$ scaling of the $M^{I=1}_{n=2,\tau=4}\left(Q^2\right)$, we require knowledge of individual operator matrix elements that contribute to $\mathcal{A}_A^{27,8}  $ and $\mathcal{A}_{B,C}^{8}  $ since the Wilson coefficients for the contributing operators (labelled \lq\lq $j$") do not have identical evolution.  The matrix elements of these operators at the input scale (see Eq.(\ref{shorthand})) are constrained by the values listed in Table~\ref{tab:mes}.  These constraints are insufficient to compute the individual values of the matrix elements appearing on the RHS of Eq.(\ref{shorthand}) however.  To do so and for purposes of illustration, we have assumed several relationships among the $\mathcal{A}^{j,(k,l)}$.  To illustrate our method, we introduce a shorthand for the four-quark operators of type A and B appearing in Eq.(\ref{shorthand})  
\bea
\mathcal{A}^{j,c}_{\pm} &=& \mathcal{A}^{j;2}\pm 2\mathcal{A}^{j;4}+\mathcal{A}^{j;6} \\
\mathcal{A}^{j}_{\pm} &=& \mathcal{A}^{j;1}\pm 2\mathcal{A}^{j;3}+\mathcal{A}^{j;5} ,
\eea
where $j$ denotes the flavor structures of the 4Q operators and takes the value A or B (see Eq.(\ref{shorthand})).  The integers $(1-6)$ specify the individual matrix element of the basis operators appearing in Eq.(\ref{jaffe-basis-1}), and $c$ denotes the presence of an SU(3) color generator.  Using the above relationships, and the computation of the $\mathcal{A}^{A,c}_{\pm}$ matrix elements from the lattice, we have the following constraints for the 4Q operators,
\bea\label{consts}
\mathcal{A}^{A;4} &=& \frac{1}{4}(\mathcal{A}^{A,c}_+-\mathcal{A}^{A,c}_-)\\
\mathcal{A}^{A;6} &=& \frac{1}{2}(\mathcal{A}^{A,c} + \mathcal{A}^{A,c}_-) - \mathcal{A}^{A;2}.
\eea  
Thus, $\mathcal{A}^{A;4}$ is determined by the lattice values for $\mathcal{A}^{A,c}_+$ and $\mathcal{A}^{A,c}_-$, while $\mathcal{A}^{A;2}$ remains unconstrained.  We take the value of $\mathcal{A}^{A;2}$ to be 25\% smaller than the matrix elements $\mathcal{A}^{A,c}_{\pm}$ and use a similar set of relations for the $\mathcal{A}^{B,(c)}_{\pm}$ matrix elements.  The above procedure is applied to all 4Q matrix elements regardless of flavor.  The values of the octet matrix elements of type A ($\mathcal{A}^{8}$ appearing in Table~\ref{tab:mes}) are taken to equal the flavor 27 matrix elements.

For the two-quark operators, the analysis in \cite{Choi:1993cu} along with Eq.(\ref{q7rel}) constrains the combination,
\bea\label{2qconst}
\mathcal{A}^{C} = \mathcal{A}^{7,k=0} - \mathcal{A}^{7,k=1}.
\eea
We have made the following {\em ansatz} to compute the individual matrix elements ($\mathcal{A}^{7,k=0,1}$),
\bea
\mathcal{A}^{7,k=0} = - \mathcal{A}^{7,k=1} = \frac{1}{2}\mathcal{A}^{C}.
\eea
This leaves the 2Q matrix elements $\mathcal{A}^{8,k=0,1}$ appearing in Eq.(\ref{jaffe-basis-1}).  We have assumed these matrix elements are equal to $\mathcal{A}^{7,k=0,1}$.  A complete Table of each matrix element, their flavor structure and the methods used to compute them can be found in Section \ref{appendixE}. 
 
%%%%%%%%%%%%%%%%%%%%%%%%%%%%%%%%%%%%%%%
%												                            %
%				DISCUSSION/PLOTS 				                            %
%											                                     %
%%%%%%%%%%%%%%%%%%%%%%%%%%%%%%%%%%%%%%%
\section{Leading Moment Analysis}\label{plots}
\begin{figure}
\begin{center}
\includegraphics[scale=0.48]{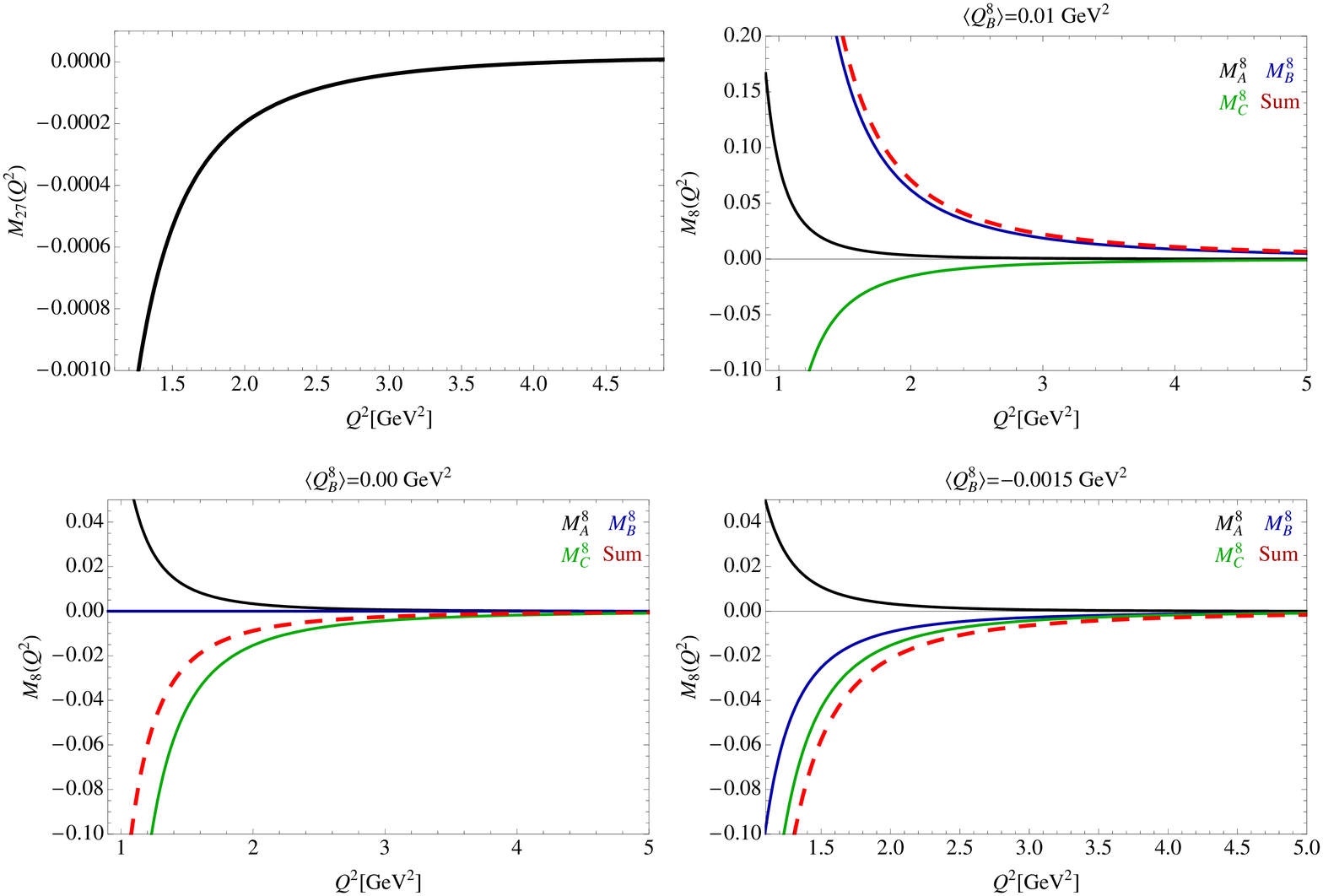}
\caption{Flavor decomposition of $M^{I=1}_{n=2,\tau=4}\left(Q^2\right)$, using the logarithmic evolution discussed in Sections \ref{wcs} and \ref{renorm} and the matrix elements in Table~\ref{totme}. The values of the $\mathcal{A}_B$ octet matrix elements are shown above each plot.  The notation $\langle Q_B^8 \rangle$ follows the notation in Table~\ref{totme}. The specific values listed above each plot are used for all type-B matrix elements contributing to the moment.  Central values were used for the two-quark octet operators for each curve.}
\label{octet}
\end{center}
\end{figure}

Using the input discussed above, we show illustrative results for the \textbf{27} and \textbf{8} contributions for values of $Q^2$ in the resonance region  in Fig.~\ref{octet}.  Under the assumptions used for the matrix elements, the twist-4 contributions to the moment are typically quite small in the resonance region. For  $\mathcal{A}_A^{8}\sim \mathcal{A}_A^{27}|_\mathrm{Latt}$, the flavor octet contribution to the moment is determined largely by the $\mathcal{A}_{B,C}$ matrix elements since $\mathcal{A}_A^{27}|_\mathrm{Latt}$ is quite small.  
%This is due to the size of the four-quark matrix element $\mathcal{A}_A^{27}$ and the ansatz that $\mathcal{A}_A^{27} = \mathcal{A}_A^{8_1}$.  
Moreover, for positive values of $\mathcal{A}_B^{8}$, cancellations occur within the octet sector, further reducing the overall octet contribution.  

\begin{figure}
\begin{center}
\includegraphics[scale=0.48]{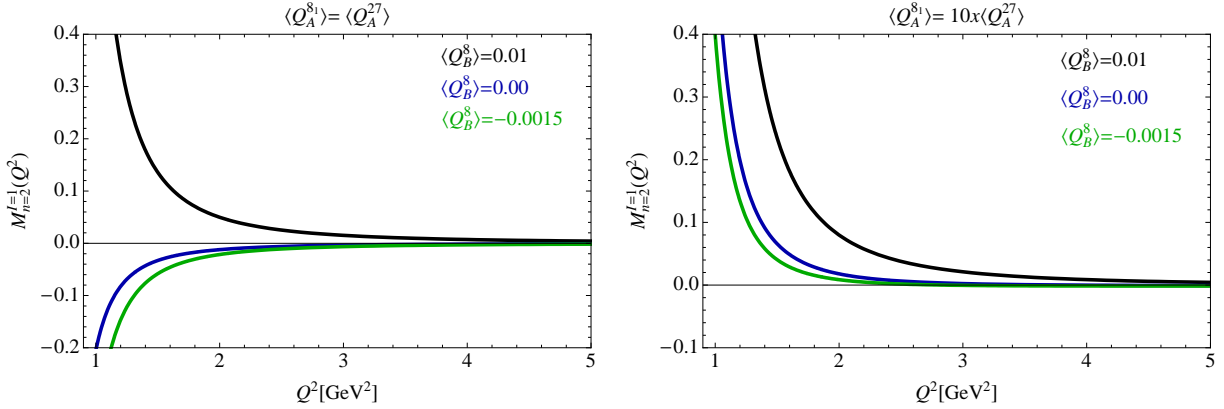}
\caption{Total contribution to $M^{I=1}_{n=2,\tau=4}\left(Q^2\right)$, see Eq.(\ref{moment}).  In the left-hand plot we have taken $\mathcal{A}_A^{8} \sim \mathcal{A}^{27}|_{\text{Latt}}$, in the right-hand plot we have taken $\mathcal{A}_A^{8}\sim 10\times\mathcal{A}_A^{27}|_\mathrm{Latt}$.  In each plot, the three curves represent different values for the reduced matrix element $\mathcal{A}_B^{8}$.  We have taken the central values appearing in Table~\ref{totme} for the $\mathcal{A}_C^{8}$ matrix elements for both plots.}
%The matrix element $\mathcal{A}_A^{27}$ was computed on the lattice (see section \ref{matrixelements}), the $\mathcal{A}_{B,C}^{8_1}$ matrix elements were estimated in Ref. \cite{Choi:1993cu}, and we have assumed $\mathcal{A}_A^{8_1} = \mathcal{A}_A^{27}$.  In the right plot, we have rescaled the y-axis to illustrate the experimental uncertainty in extracting the leading moment at twist-4 \cite{Osipenko:2003bu}.}
\label{RGE}
\end{center}
\end{figure}

The total contribution to $M^{I=1}_{n=2,\tau=4}\left(Q^2\right)$ is given in Fig.~\ref{RGE}.  We plot two representative cases for differing values of $\mathcal{A}_A^{8}$.  In the left-hand plot, we have taken $\mathcal{A}_A^{8} \sim \mathcal{A}_A^{27}|_\mathrm{Latt}$.  In the event that the behaviors of the leading proton and isovector moments are similar,  this value would be disfavored by the CLAS data.  CLAS has measured the twist four moment to be positive (Ref~\cite{Osipenko:2003bu}) in the range of $Q^2$ shown in Fig.~\ref{RGE}.     In the right-hand plot, we have taken $\mathcal{A}_A^{8} \sim 10\times\mathcal{A}_A^{27}|_\mathrm{Latt}$.  Here, a range of small positive values of $\mathcal{A}^{8}_B$ yields an isovector moment which is qualitatively similar to the CLAS data for the proton.  It is possible, then, that the CLAS data provide a hint of a hierarchy among the twist-four operator matrix elements, though a definitive statement will require data on $F_2$ for the neutron as well as an improvement in overall experimental precision.  At present, the experimental error on the determination of the twist-four contribution to the leading moment of the proton $F_2$ structure function is  $\pm 0.015$  for $Q^2\approx 1\>\text{GeV}^2$ \cite{Osipenko:2003bu,Melnitchouk:2005zr}, while the magnitude of the leading twist contribution to the isovector moment is $\approx 0.1$ throughout the indicated region. The systematic error in obtaining the moments from data consists of genuine uncertainties in the data, as well as uncertainties in the evaluation procedure.

For comparison to the theory predictions, we note that the CLAS data for $F_2$ is extracted from electron-proton DIS and thus is not the linear combination of moments shown in Eqs.(\ref{flavormoment1},\ref{flavormoment2}).  The experimental data are also not separated into flavor singlet and non-singlet channels, and thus include the effects of the gluonic operators shown in Eq.(\ref{jaffe-basis-2}). 
% In Fig.~[\ref{RGE}], the twist-2 pieces of the moment have been subtracted out so that only the twist-4 pieces remain.   
One can see from Fig.~\ref{RGE} that the experimental precision in extracting the leading moment for higher twist does not yet allow one to disentangle the separate flavor channels computed here.  Next generation experiments may be required to probe the higher twist flavor contributions to the leading moment of $F_2$.  

\begin{figure}
\begin{center}
\includegraphics[scale=0.37]{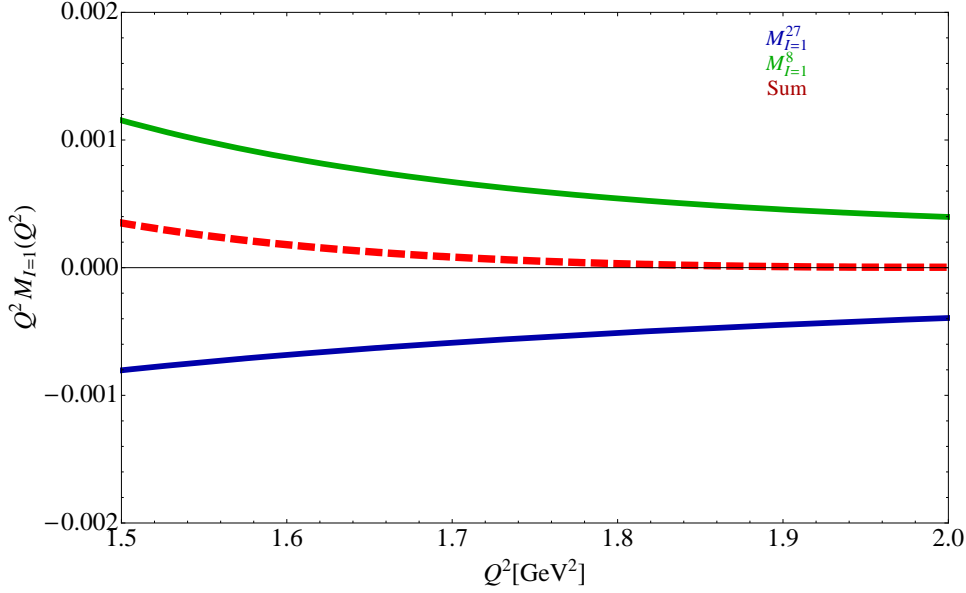}
\caption{We have tuned the values of $\mathcal{A}_B^{8}$ and $\mathcal{A}_C^{8}$ to cancel against the flavor \textbf{27} moment at $Q^2\approx5\>\text{GeV}^2$.  The chosen values for the reduced matrix elements, shown in Eq.(\ref{values}), are consistent with Table~\ref{tme}.  We have multiplied all curves by $Q^2$ to illustrate the breaking of this cancellation is due to QCD evolution. }
\label{canc}
\end{center}
\end{figure}

 Finally, we explore the possibility that the twist-4 contribution may be suppressed at one scale due to cancellations between operator contributions and that $Q^2$-evolution may lead to a breakdown of this cancellation at other scales.
In Fig.~\ref{canc}, we have tuned the values of $\mathcal{A}_B^{8}$ and $\mathcal{A}_C^{8}$ such that the total octet contribution to the isovector moment cancels the flavor \textbf{27} contribution at $Q^2\approx 2\>$GeV$^2$.  The total moment is given by the dashed curve in Fig.~\ref{canc}.  The chosen values of the reduced matrix elements are consistent with Table~\ref{tme} and are given by,
\bea\label{values}
\mathcal{A}_{A}^{27}&=&(25,\>-1.3,\>25,\>-1.0,\>25,\>-1.8)^T\times 10^{-4}\>\mathrm{GeV}^2\\
\mathcal{A}_{A}^{8}&=&(25,\>-1.3,\>25,\>-1.0,\>25,\>-1.8)^T\times 10^{-4}\>\mathrm{GeV}^2\\
\mathcal{A}_{B}^{8}&=&(1.96,\>1.96,\>1.96,\>1.96,\>1.96,\>1.96)^T\times 10^{-3}\>\mathrm{GeV}^2\\
\mathcal{A}_{C}^{8}&=&(-0.035,\> 0.035,\> -0.035, \>0.035)^T\>\mathrm{GeV}^2.
\eea
   The sum of the \textbf{27} and octet moment is given by the dashed curve in Fig.~\ref{canc} which vanishes at the input scale of $Q^2=5\>\text{GeV}^2$.  At smaller values of $Q^2$, the dashed curve deviates from zero.  We have multiplied the moment by $Q^2$ so that one can see more clearly that the spoiling of this cancellation is due to QCD evolution.  The magnitude of the departure from this cancellation is well below the present CLAS experimental error, so an observation of this effect -- should it be realized in nature -- would again require significant improvements in experimental precision, at least for the leading moment.

%Assuming no further cancellations occur between the iso-singlet contributions to the moment and the octet contributions, the value $\langle\bar{\psi}Q^2\psi \bar{\psi}\psi\rangle = -0.03\>\text{GeV}^2$ is disfavored by experiment for the leading moment, (for details, see \cite{Osipenko:2003bu}).    
%Separate flavor contributions to the leading moment at twist-4.  The top row of plots correspond to the \textbf{27} representation whereas the bottom row corresponds to both the $\textbf{8}_1$ and $\tilde{\textbf{8}}_1$ representations.  The power dependence ($1/Q^2$) has been removed from the left-hand column of plots by multiplying the moment by $Q^2$, the right-hand column includes the $1/Q^2$ effects.  The grayed regions in Fig.[\ref{flavorseparate}] represent the theoretical uncertainty in the moment, taken from the lattice uncertainty in computing the four-quark reduced matrix element.  We point out here that the $Q^2$-power law dependence does not play a dramatic role in the evolution of the moment except at very low $Q^2$.  The twist-4 contributions to the moment are quite small in the resonance region.

%%%%%%%%%%%%%%%%%%%%%%%%%%%%%%%%%%%%%%%
%												                            %
%				CONCLUSIONS 	 				                            %
%											                                     %
%%%%%%%%%%%%%%%%%%%%%%%%%%%%%%%%%%%%%%%
\section{Conclusions and Outlook}\label{disc}
In this work, we have provided the first complete computation of the logarithmic evolution of twist-four contributions to the flavor non-singlet, leading moment of the structure function $F_2$. Our results can be employed to analyze future deep-inelastic scattering experiments carried out in the resonance regime, going beyond the simple {\em ansatz} given in Eq.~(\ref{param}) and utilized in Refs.~~\cite{Osipenko:2003bu,Ricco:1998yr, Simula:2000ka}. In the present instance, we have used our calculation to carry out an illustrative phenomenological study the magnitude and $Q^2$-dependence of the leading moment, drawing on existing lattice computations and phenomenological determinations of a subset of the operator matrix elements and plausible {\em ansatz} for the others. Despite lacking a complete set of non-perturbative QCD matrix element computations, we are able to draw a few broad conclusions from our study:
\begin{itemize}
\item Theoretically one expects the overall scale of twist-4 contributions to the leading, flavor non-singlet moment to be small and consistent in magnitude with the results obtained from the CLAS analysis\cite{Osipenko:2003bu}. 
\item The suppressed scale may result either from all matrix elements individually being small or a cancellation between various twist-4 contributions.
\item The CLAS data may provide early hints of a hierarchy among operator matrix elements, possibly indicating a larger magnitude for the flavor \textbf{8} than for the flavor \textbf{27}
\item Additional data on $F_2$ for the neutron as well as improvement in overall experimental precision will be needed to further disentangle contributions from different flavor channels as well as to observe a cancellation scenario if it exists.
\item A full QCD prediction for the leading moment requires computation of several non-perturbative matrix elements.
\end{itemize}

To amplify on these remarks, we note that the opposite sign behavior in Fig.~\ref{octet} allows for cancellation effects within twist-4.  Such cancellations have been alluded to as the reason the higher twist moments are essentially independent of $Q^2$ in the resonance region \cite{Osipenko:2003bu}, leading to a scaling behavior in the moment of $F_2(x,Q^2)$ known as Bloom-Gilman duality.  Within past analyses, however, cancellations are typically found between the twist-4 and higher twist pieces in the twist expansion.  Here we find evidence that even within twist-4, there is a possibility of cancellation independent of the higher twist matrix elements being small.  However, we note that the twist-4 contribution to the moment is quite small over a wide range of $Q^2$.  This suggests that the cancellation effects within twist-4 may not in themselves be the cause of the twist-4 moment being so modest and that small size of the individual matrix elements is also responsible.
%It is well known that the elastic contributions to the moment begins to dominate in the lowest regions of $Q^2$ in Fig.~\ref{RGE}, see Ref.\cite{Armstrong:2001xj}.  As pointed out there, an analysis of the data in terms of the OPE in this region is hampered by the fact that the low $n$ moments are mainly sensitive to the inclusion of elastic contributions, clouding interpretations in terms of higher twist effects.  

This issue can be viewed as being due to the relative sizes of the matrix elements.  We note that the scale of both the \textbf{27} and \textbf{8} moments is set by the matrix elements of the twist-4 operators, and given different, possibly larger matrix elements, one may see adequate variation in the moment at a higher scale in $Q^2$.  The current experimental precision in extracting the leading moment for higher twist does not allow a determination of the separate flavor channels computed here.  Thus probing cancellation effects would require higher level of experimental precision.  From the theoretical perspective, one need not stop at a leading moment analysis however. 

Higher twist effects are expected to play a larger role at a given $Q^2$ for larger moment.  Heuristically this is due to the fact that the resonance region weights the large-$x$ region more heavily and is thus dominated by larger-$n$ moments.   This effect is clearly illustrated in the CLAS data for $ep$ scattering where the moment is decomposed in terms of the parametrization given in Eq.(\ref{param}).  It is thus desirable to compute the anomalous dimension for the operators appearing in Eq.(\ref{jaffe-basis-1}) for arbitrary $n$ since experiments are more sensitive to higher twist effects for larger $n$.  The technical challenges encountered in going beyond the leading moment are quite demanding however, both from perturbative QCD viewpoint and also from a lattice QCD perspective.  One of the main advantages in performing a leading moment analysis is due to the already existing lattice estimates of the tree level 4Q operators, and we know of no lattice computation of these matrix elements beyond leading spin that exists in the literature.  Due to these technical challenges, we leave a higher spin analysis of the type presented in here for future work.  
\newpage

\noindent \noindent{\it  Acknowledgements}:  The authors thank A. Belitksy, H. Patel, and P. McGuirk  for valuable discussions and helpful comments during the course of this work. The research is partially supported by U.S. Department of Energy contracts  DE-FG02-08ER41531 (MJRM) and the Wisconsin Alumni Research Foundation ( MJRM), the U.S. National Science Foundation under grant NSF-PHY- 0705682 (SM), and the U.S. Department of Energy under grant DE-FG05-84ER40154 (MJG).

%%%%%%%%%%%%%%%%%%%%%%%%%%%%%%%%%%%%%%%
%												                            %
%				APPENDIX A						                            %
%											                                     %
%%%%%%%%%%%%%%%%%%%%%%%%%%%%%%%%%%%%%%%
\appendix
\section{Anomalous Dimension Matrices}\label{apxA}

We give here the anomalous dimension matrices for operators introduced in section \ref{sec-flavor}.  When computing the one loop corrections to these operators, each operator was inserted into a Green's function with the same number of external legs as the operator itself, and dimensional regularization was employed where $d=4-2\epsilon$.  For the 2Q Feynman diagrams shown in Fig.~\ref{2Qrenorm}, we have calculated each using the background field method and have chosen the Feynman gauge for all calculations \cite{Abbott:1980hw}.    Incorporating operator mixing, the anomalous dimension is, in general, a matrix in flavor space

Following the notation in section \ref{anom-dim-flavor} the 27-plet anomalous dimension matrix is
%---------------------------------------------------------------------%
\bea
\left(
\begin{array}{c}
 \vec{O}^{27}_{I=2} \\
 \vec{O}^{27}_{I=1} \\
\vec{O}^{27}_{I=0} \\

\end{array}
\right)_b=\left(
\begin{array}{cccc}
\mathbb{P}_{I=2} & 0 & 0   \\
 0 & \mathbb{P}_{I=1} & 0   \\
 0 & 0 & \mathbb{P}_{I=0} \\
\end{array}
\right)\left(
\begin{array}{c}
 \vec{O}^{27}_{I=2} \\
 \vec{O}^{27}_{I=1} \\
 \vec{O}^{27}_{I=0} \\
 \end{array}
\right),
\eea
%---------------------------------------------------------------------%
\begin{equation}
\mathbb{P}_{\text{I}=2,1,0} = \frac{g^2}{4\pi^2} \left(
\begin{array}{cccccc}
 \frac{11}{2} & \frac{35}{6} & 0 & 0 & 0 & 0 \\
 \frac{119}{18} & \frac{35}{6} & 0 & 0 & 0 & 0 \\
 0 & 0 & \frac{11}{2} & \frac{47}{6} & 0 & 0 \\
 0 & 0 & \frac{127}{18} & \frac{20}{3} & 0 & 0 \\
 0 & 0 & 0 & 0 & \frac{11}{2} & \frac{35}{6} \\
 0 & 0 & 0 & 0 & \frac{119}{18} & \frac{35}{6}
\end{array}
\right)
\end{equation}
%---------------------------------------------------------------------%
The anomalous dimension matrix for the octet sector has the schematic form
%---------------------------------------------------------------------%
\bea
\left(
\begin{array}{c}
 \vec{O}^{8,A}_{I=1} \medskip \\
 \vec{O}^{8,B,C}_{I=1} 
 \end{array}
\right)_b=\left(
\begin{array}{cc}
  \mathbb{L}_{I=1} & \mathbb{M}_{I=1} \medskip\\
 \mathbb{Q}_{I=1} & \mathbb{N}_{I=1} \\
 \end{array}
\right)\left(
\begin{array}{c}
 \vec{O}^{8,A}_{I=1} \medskip\\
 \vec{O}^{8,B,C}_{I=1} \\
 \end{array}
\right).
\eea
%---------------------------------------------------------------------%
And each of $\mathbb{L}, \mathbb{M},\mathbb{Q},\mathbb{N}$ are divided according to the mixings between two and 4Q operators
\bea
\mathbb{L}_{I}&=&\left(
\begin{array}{cc}
\mathbb{L}_{I}^{4Q\rightarrow 4Q} &\mathbb{L}_{I}^{4Q\rightarrow 2Q}  \\
\mathbb{L}_{I}^{2Q\rightarrow 4Q} & \mathbb{L}_{I}^{2Q\rightarrow 2Q}  \\
\end{array}
\right)
\qquad
\mathbb{N}=
\left(
\begin{array}{cc}
\mathbb{N}_{I}^{4Q\rightarrow 4Q} & \mathbb{N}_{I}^{4Q\rightarrow 2Q}  \\
\mathbb{N}_{I}^{2Q\rightarrow 4Q} & \mathbb{N}_{I}^{2Q\rightarrow 2Q} \\
\end{array}
\right)\nn \\
%---------------------------------------------------------------------%
\mathbb{M}_{I}&=&\left(
\begin{array}{cc}
\mathbb{M}_{I}^{4Q\rightarrow 4Q} & \mathbb{M}_{I}^{4Q\rightarrow 2Q}  \\
\mathbb{M}_{I}^{2Q\rightarrow 4Q}  & \mathbb{M}_{I}^{2Q\rightarrow 2Q} \\
\end{array}
\right)
\qquad
\mathbb{Q}_{I}=\left(
\begin{array}{cc}
\mathbb{Q}_{I}^{4Q\rightarrow 4Q} & \mathbb{Q}_{I}^{4Q\rightarrow 2Q}  \\
\mathbb{Q}_{I}^{2Q\rightarrow 4Q}  & \mathbb{Q}_{I}^{2Q\rightarrow 2Q} \\
\end{array}
\right)
\eea
%---------------------------------------------------------------------%
For $\mathbb{L}_{I=1}$, the only non-zero sector is the four-quark mixing back to four-quark piece.
%---------------------------------------------------------------------%
\bea
\mathbb{L}_{I=1}=\left(
\begin{array}{cc}
\mathbb{L}_{I=1}^{4Q\rightarrow 4Q} & \mathbf{0}  \\
\mathbf{0} & \mathbf{0} \\
\end{array}
\right)
\qquad\text{where}\qquad
\mathbb{L}^{4Q\rightarrow 4Q}_{I=1} = \frac{g^2}{4\pi^2} \left(
\begin{array}{cccccc}
 \frac{11}{2} & \frac{35}{6} & 0 & 0 & 0 & 0 \\
 \frac{119}{18} & \frac{35}{6} & 0 & 0 & 0 & 0 \\
 0 & 0 & \frac{11}{2} & \frac{47}{6} & 0 & 0 \\
 0 & 0 & \frac{127}{18} & \frac{20}{3} & 0 & 0 \\
 0 & 0 & 0 & 0 & \frac{11}{2} & \frac{35}{6} \\
 0 & 0 & 0 & 0 & \frac{119}{18} & \frac{35}{6}
\end{array}
\right)\eea
%---------------------------------------------------------------------%
For $\mathbb{M}_{I=1}$, we find its anomalous dimension to be
%---------------------------------------------------------------------%
\bea
\mathbb{M}_{I=1}=\left(
\begin{array}{cc}
\mathbb{M}_{I=1}^{4Q\rightarrow 4Q} & \mathbf{0}  \\
\mathbf{0} & \mathbf{0} \\
\end{array}
\right)
\qquad\text{where}\qquad
\mathbb{M}^{4Q\rightarrow 4Q}_{I=1}=\frac{g^2}{4\pi^2}\left(
\begin{array}{cccccc}
 0 & \frac{17}{2} & 0 & \frac{17}{2} & 0 & 0 \\
 0 & \frac{64}{9} & 0 & \frac{64}{9} & 0 & 0 \\
 0 & 0 & 0 & 0 & 0 & 0 \\
 0 & 0 & 0 & 0 & 0 & 0 \\
 0 & 0 & 0 & \frac{17}{2} & 0 & \frac{17}{2} \\
 0 & 0 & 0 & \frac{64}{9} & 0 & \frac{64}{9}
\end{array}
\right)
\eea
%---------------------------------------------------------------------%
The anomalous dimension for $\mathbb{Q}_{I=1,0}$ vanishes.  For $\mathbb{N}_{I=1}$ the anomalous dimension has a schematic form 
%---------------------------------------------------------------------%
\bea
\mathbb{N}_{I=1}=
\left(
\begin{array}{cc}
\mathbb{N}_{I=1}^{4Q\rightarrow 4Q} & \mathbf{0}  \\
\mathbb{N}_{I=1}^{2Q\rightarrow 4Q} & \mathbb{N}_{I=1}^{2Q\rightarrow 2Q} \\
\end{array}
\right)_\mathbf{10x10}
\eea
Where the non-vanishing sectors of this matrix are,
%---------------------------------------------------------------------%
\bea
\mathbb{N}^{4Q\rightarrow 4Q}_{I=1}=\frac{g^2}{4\pi^2}
\left(
\begin{array}{cccccc}
 \frac{11}{2} & \frac{13}{2} & 0 & \frac{15}{2} & 0 & 0 \\
 \frac{119}{18} & \frac{58}{9} & 0 & \frac{67}{9} & 0 & 0 \\
 0 & 0 & \frac{11}{2} & \frac{47}{6} & 0 & 0 \\
 0 & \frac{43}{6} & \frac{127}{18} & \frac{43}{6} & 0 & 7 \\
 0 & 0 & 0 & \frac{15}{2} & \frac{11}{2} & \frac{13}{2} \\
 0 & 0 & 0 & \frac{67}{9} & \frac{119}{18} & \frac{58}{9}
\end{array}
\right)
\qquad
\mathbb{N}^{2Q\rightarrow 2Q}_{I=1}=\frac{g^2}{4\pi^2}
\left(
\begin{array}{cccc}
 \frac{1151}{144} & \frac{415}{144} & \frac{19}{8} & -\frac{25}{24} \\
 \frac{193}{72} & \frac{187}{24} & \frac{31}{9} & -\frac{16}{9} \\
 \frac{89}{144} & \frac{121}{144} & \frac{587}{72} & \frac{145}{72} \\
 \frac{29}{96} & \frac{23}{288} & \frac{59}{48} & \frac{107}{16}
\end{array}
\right)\nn
\eea

\bea
\mathbb{N}^{2Q\rightarrow 4Q}_{I=1}=\frac{g^2}{4\pi^2}
\left(
\begin{array}{cccccc}
 \frac{16}{27} & \frac{61}{72} & -\frac{8}{27} & \frac{9}{4} & -\frac{8}{27} & \frac{181}{72} \\
 \frac{8}{27} & -\frac{37}{36} & \frac{8}{27} & \frac{13}{18} & -\frac{16}{27} & \frac{23}{36} \\
 -\frac{1}{27} & 0 & -\frac{2}{27} & \frac{1}{2} & -\frac{1}{27} & 0 \\
 -\frac{1}{27} & -\frac{41}{144} & \frac{2}{9} & -\frac{7}{24} & -\frac{1}{27} & -\frac{41}{144}
\end{array}
\right)
\eea
%---------------------------------------------------------------------%
The matrix $\mathbb{L}_{I=0}$ has only one, non-vanishing piece
%---------------------------------------------------------------------%
\bea
\mathbb{L}_{I=0}=\left(
\begin{array}{cc}
\mathbb{L}_{I=0}^{4Q\rightarrow 4Q} & \mathbf{0}  \\
\mathbf{0} & \mathbf{0} \\
\end{array}
\right)
\qquad \text{where}\qquad
\mathbb{L}^{4Q\rightarrow 4Q}_{I=0}=\frac{g^2}{4\pi^2} \left(
\begin{array}{cccccc}
 \frac{11}{2} & \frac{35}{6} & 0 & 0 & 0 & 0 \\
 \frac{119}{18} & \frac{35}{6} & 0 & 0 & 0 & 0 \\
 0 & 0 & \frac{11}{2} & \frac{47}{6} & 0 & 0 \\
 0 & 0 & \frac{127}{18} & \frac{20}{3} & 0 & 0 \\
 0 & 0 & 0 & 0 & \frac{11}{2} & \frac{35}{6} \\
 0 & 0 & 0 & 0 & \frac{119}{18} & \frac{35}{6}
\end{array}
\right)
\eea
%---------------------------------------------------------------------%
Similarly $\mathbb{M}_{I=0}$ has only one sector which does not vanish, it is 
%---------------------------------------------------------------------%
\bea
\mathbb{M}_{I=0}=\left(
\begin{array}{cc}
\mathbb{M}_{I=0}^{4Q\rightarrow 4Q} & \mathbf{0}  \\
\mathbf{0} & \mathbf{0} \\
\end{array}
\right)
\qquad \text{where}\qquad
\mathbb{M}^{4Q\rightarrow 4Q}_{I=0}=\frac{g^2}{4\pi^2}\left(
\begin{array}{cccccc}
 0 & \frac{17}{2} & 0 & \frac{17}{2} & 0 & 0 \\
 0 & \frac{64}{9} & 0 & \frac{64}{9} & 0 & 0 \\
 0 & 0 & 0 & 0 & 0 & 0 \\
 0 & 0 & 0 & 0 & 0 & 0 \\
 0 & 0 & 0 & \frac{17}{2} & 0 & \frac{17}{2} \\
 0 & 0 & 0 & \frac{64}{9} & 0 & \frac{64}{9}
\end{array}
\right)
\eea
%---------------------------------------------------------------------%
Finally, this leaves $\mathbb{N}_{I=0}$
%---------------------------------------------------------------------%
\bea
\mathbb{N}_{I=0}=
\left(
\begin{array}{cc}
\mathbb{N}_{I=0}^{4Q\rightarrow 4Q} & \mathbf{0}  \\
\mathbb{N}_{I=0}^{2Q\rightarrow 4Q} & \mathbb{N}_{I=0}^{2Q\rightarrow 2Q} \\
\end{array}
\right)_\mathbf{10x10}
\eea
the three non-vanishing sectors of $\mathbb{N}_{I=0}$ are,
%---------------------------------------------------------------------%
\bea
\mathbb{N}^{4Q\rightarrow 4Q}_{I=0}=\frac{g^2}{4\pi^2}
\left(
\begin{array}{cccccc}
 \frac{11}{2} & \frac{13}{2} & 0 & \frac{15}{2} & 0 & 0 \\
 \frac{119}{18} & \frac{58}{9} & 0 & \frac{67}{9} & 0 & 0 \\
 0 & 0 & \frac{11}{2} & \frac{47}{6} & 0 & 0 \\
 0 & 0 & \frac{127}{18} & \frac{43}{6} & 0 & \frac{85}{12} \\
 0 & 0 & 0 & \frac{15}{2} & \frac{11}{2} & \frac{13}{2} \\
 0 & 0 & 0 & \frac{67}{9} & \frac{119}{18} & \frac{58}{9}
\end{array}
\right)
\qquad
\mathbb{N}^{2Q\rightarrow 2Q}_{I=0}=\frac{g^2}{4\pi^2}
\left(
\begin{array}{cccc}
 \frac{1151}{144} & \frac{415}{144} & \frac{19}{8} & -\frac{25}{24} \\
 \frac{193}{72} & \frac{187}{24} & \frac{31}{9} & -\frac{16}{9} \\
 \frac{89}{144} & \frac{121}{144} & \frac{587}{72} & \frac{145}{72} \\
 \frac{29}{96} & \frac{23}{288} & \frac{59}{48} & \frac{107}{16}
\end{array}
\right)
\eea
%---------------------------------------------------------------------%
\bea
\mathbb{N}^{2Q\rightarrow 4Q}_{I=0}=\frac{g^2}{4\pi^2}
\left(
\begin{array}{cccccc}
 \frac{16}{27} & \frac{61}{72} & -\frac{8}{27} & \frac{9}{4} & -\frac{8}{27} & \frac{181}{72} \\
 \frac{8}{27} & -\frac{37}{36} & \frac{8}{27} & \frac{13}{18} & -\frac{16}{27} & \frac{23}{36} \\
 -\frac{1}{27} & 0 & -\frac{2}{27} & \frac{1}{2} & -\frac{1}{27} & 0 \\
 -\frac{1}{27} & -\frac{41}{144} & \frac{2}{9} & -\frac{7}{24} & -\frac{1}{27} & -\frac{41}{144}
\end{array}
\right)
\eea

%%%%%%%%%%%%%%%%%%%%%%%%%%%%%%%%%%%%%%%
%												                            %
%				APPENDIX B						                            %
%											                                     %
%%%%%%%%%%%%%%%%%%%%%%%%%%%%%%%%%%%%%%%
\section{Wilson Coefficients}\label{apxB}
We list here the full expressions for the Wilson coefficients derived in \cite{Jaffe:1981td}.  Starting with the forward Compton amplitude,  
%\bea
%T_{\mu\nu} &=& -i \int \frac{d^4 x}{(2\pi)} \>e^{iq \cdot x } \> \langle \text{N} | \> \text{T}\left(J_\mu(x)J_\nu(0)\right) \> | \text{N} \rangle   \nn\\
%&=& -i(q_\mu q_\nu-q^2 g_{\mu\nu})\sum_{i,n} \langle \text{N} | \mathcal{O}^{(i)}_{L,\mu_1\ldots \mu_n} | \text{N} \rangle  \left\{ \int d^4x \ \ x^{\mu_1}  \ldots x^{\mu_n} C^{(i)}_{L,n} \ \ e^{i q\cdot x}\right\} \nn\\
%&-&i (g_{\mu\lambda} q_\rho q_\nu+g_{\rho \nu}q_\mu q_\lambda-q^2 g_{\mu\lambda}g_{\rho\nu}-g_{\mu\nu}q_\lambda q_\rho)\sum_{i,n} \langle \text{N} |\mathcal{O}^{(i),\lambda\rho}_{2,\mu_1\ldots \mu_n} |\text{N} \rangle  \left\{ \int d^4x \ \ x^{\mu_1}  \ldots x^{\mu_n} C^{(i)}_{2,n} \ \ e^{i q\cdot x}\right\} \nn\\
%\eea
\bea
T_{\mu\nu} &=& -i \int \mathrm{d}^4x \>e^{iq\cdot x} \langle \mathrm{N} | \mathrm{T}\left(J_\mu(x)J_\nu(0)\right) | \mathrm{N}\rangle \nn\\
\nn\\
&=& \sum_{i,n}\left(\frac{2}{-q^2}\right)^n\left[\left(g_{\mu\nu}-\frac{q_\mu q_\nu}{q^2}\right)q_{\mu_1}\ldots q_{\mu_n}C_{L,i}^{n}\mathcal{O}^{i}_{L,\mu_1\ldots\mu_n} \right. \nn\\
&&\left.-(g_{\mu\mu_1}g_{\nu\mu_2}q^2-g_{\mu\mu_1}q_\nu q_{\mu_2}-g_{\nu\mu_2}q_{\mu}q_{\mu_1}+g_{\mu\nu}q_{\mu_1}q_{\mu_2})q_{\mu_3}\ldots q_{\mu_n} C_{2,i}^n\mathcal{O}^{i}_{2,\mu_1\ldots\mu_n}\right] \nn\\
\nn\\
&=&\sum_{i,n}\left[\left(g_{\mu\nu}-\frac{q_\mu q_\nu}{q^2}\right)\omega^n C_{L,i}^n \mathcal{A}^i_{L,n}\right.\nn\\
&&\left.-\left(\frac{2}{-q^2}\right)^n\left\{p_\mu p_\nu q^2 - p_\mu q_\nu q\cdot p - p_\nu q_\mu q\cdot p + g_{\mu\nu} (q\cdot p)^2 \right\} (q\cdot p)^{n-2}C_{2,i}^n\mathcal{A}^i_{2,n} \right]  
\nn\\
\nn\\
&=& \sum_{i,n} \left[e_{\mu\nu} C_{L,i}^n\mathcal{A}^i_{L,n} + d_{\mu\nu}C_{2,i}^n\mathcal{A}^i_{2,n}\right]\omega^n.
\eea
In the second line above, we have parametrized the symmetric and traceless matrix elements in terms of the nucleon four-momenta,
\begin{eqnarray*}
\begin{cases}
\langle \text{N} |\mathcal{O}^{(i),\tau=4}_{L,\mu_1\ldots\mu_n} |\text{N}  \rangle \ &= \mathcal{A}^{(i),\tau=4}_{L,n}\left( p_{\mu_1} \ldots p_{\mu_n} - \text{traces}\right)  \\
\\
\langle \text{N} | \mathcal{O}^{(i)\tau=4}_{2,\mu_1\ldots\mu_n} |\text{N}  \rangle &= \mathcal{A}^{(i),\tau=4}_{2,n}\left( p_{\mu_1} \ldots p_{\mu_n}- \text{traces}\right). 
\end{cases}
\end{eqnarray*}
%
%\begin{eqnarray*}
%\begin{cases}
%\int \text{d}^4x \> x^{\mu_1}\ldots x^{\mu_n} \>C^{(i)}_{2,n}\> e^{iq\cdot x} \> &= \> \tilde{C}^{(i)}_{2,n+2}\left\{(-2i) \left(-\frac{2}{q^2}\right)^{n+2}q^{\mu_1}\ldots q^{\mu_n}\right\} \\
%\\
%\int \text{d}^4x \> x^{\mu_1}\ldots x^{\mu_n} \>C^{(i)}_{L,n}\> e^{iq\cdot x} \> &= \> \tilde{C}^{(i)}_{L,n}\left\{(-i) \left(-\frac{2}{q^2}\right)^{n+1}q^{\mu_1}\ldots q^{\mu_n}\right\}
%\end{cases}
%\end{eqnarray*}
%Plugging these results in for the time-ordered product of currents and simplifying, we arrive at the schematic form of the time-ordered product 
%
%\bea
%T_{\mu\nu} = (-2) \sum_{i,n} \left\{e_{\mu\nu}\mathcal{A}^{(i)}_{L,n}\tilde{C}^{(i)}_{L,n}+d_{\mu\nu}\mathcal{A}^{(i)}_{2,n}\tilde{C}^{(i)}_{2,n}\right \}\omega^n
%\eea
The sum over n is even, $i$ distinguishes the type of operator, and $n$ denotes the spin of the operator, we have also used the shorthand expressions,
\bea
e_{\mu\nu} &=& g_{\mu\nu} -\frac{q_\mu q_\nu}{q^2}\\
d_{\mu\nu} &=& -g_{\mu\nu}+\frac{(p_\mu q_\nu+q_\nu p_\nu)}{(p\cdot q)}-\frac{q^2 \> p_\mu p_\nu }{(p\cdot q)^2}  \\
\omega &=& -\frac{2 \> p\cdot q}{q^2}.
\eea
The coefficients of the tensor $e_{\mu\nu}$ contribute to $F_L$ and whereas the coefficients of $d_{\mu\nu}$ contribute to $F_2$.  The full expressions for $Y_{\mu\nu}$ and $X_{\mu\nu}$ are \cite{Jaffe:1982pm}
\begin{small}
\bea\label{ylarge}
Y_{\mu\nu}^{T=4} &=& - \frac{g}{q^6} \sum^{\infty}_{n=2_{(\text{even})}}\left (\frac{2}{q^2}\right )^{n-2} T_{\mu\nu}^{\mu_1\mu_2} \ \ q^{\mu_3 }\ldots q^{\mu_n} \nn \\
&\times& \sum_{k=0}^{n-2} \sum_{l=0}^{n-2-k}\left \{ \mathcal{O} ^{1(k,l)}_{n,\mu_1 \ldots \mu_n} \left [ \frac{n!}{k! l! (n-1-k-l)!}\left [ \frac{1}{n-k}-\frac{1}{n-l}\right ] 
+  (-1)^{k+l}\frac{(l+k+1)!}{k! l!} \left[ \frac{1}{k+1}-\frac{1}{l+1}\right ]\right ]\right . \nn \\
&+& \left. \mathcal{O}^{2 (k,l)}_{n,\mu_1 \ldots \mu_n}\left [ \frac{n!}{k! l! (n-1-k-l)!}\left[ \frac{1}{n-k}+\frac{1}{n-l}\right] + (-1)^{k+l}\frac{(l+k+1)}{k! l!} \left[ \frac{1}{k+1}+\frac{1}{l+1} \right ] \right] \right \} \nn \\
\eea
\end{small}
\begin{small}
\bea\label{xlarge}
X_{\mu\nu}^{T=4} & =& g \sum^{\infty}_{n=2 _{\text{even}}} \left[ -\frac{2}{q^2}\right ]^{n+1}\frac{1}{(n+2)^2} \nn \\
&\times& \left\{ \left[ \frac{q^\mu q^\nu}{q^2} - g^{\mu\nu}\right] \left [ -(n+1)\sum_{k=0}^{n-1}q \cdot \mathcal{O}_{n}^{3(k)} -2(n+1)\sum_{k=0}^{n-3}\sum_{l=0}^{n-3-k}(l+1)q\cdot \mathcal{O}_n^{5(k,l)}\right . \right. \nn\\
&-& \left . \left .4 \sum^{n-3}_{k=0} \sum^{n-3-k}_{l=0} (k+1)(n-2-k-l) q\cdot \mathcal{O}_n^{5(k,l)}+2(n+1)\sum_{k=0}^{n-3}\sum_{l=0}^{n-3-k}(l+1)q \cdot \mathcal{O}_n^{6(k,l)} \right. \right . \nn \\
 &+& \left . \left. 2 \sum^{n-2}_{k=0}(-1)^{k}(k+1)(n-1-k)q\cdot \mathcal{O}_n^{k} \right ] +\left [ g^{\mu\nu} - \frac{p^\mu q^\nu+p^\nu q^\mu}{p \cdot q} + \frac{q^2 p^\mu p^\nu}{(p\cdot q)^2} \right ] \right .\nn \\
&\times& \left . \left [ \frac{n(n-1)}{4} \sum^{n-1}_{k=0} q\cdot \mathcal{O}^{3(k)}_n - \frac{(5n +4)}{2} \sum^{n-3}_{k=0}\sum^{n-3-k}_{l=0}(l+1)q\cdot \mathcal{O}^{5(k,l)}_{n}\right . \right . \nn \\
&-& \left . \left . (n+8) \sum^{n-3}_{k=0}\sum^{n-3-k}_{l=0}(k+1)(n-2-k-l)q\cdot \mathcal{O}_n^{5(k,l)}+\frac{(5n+4)}{2}\sum^{n-3}_{k=0} \sum^{n-3-k}_{l=0}(l+1) q\cdot \mathcal{O}_n^{6(k,l)} \right . \right. \nn \\
&+& \left . \left . \frac{(n+8)}{2} \sum_{k=0}^{n-2} (-1)^{k}(k+1)(n-1-k)q\cdot \mathcal{O}^{7(k)}_{n}\right ] \right \}
\eea
\end{small}
Where 
\bea
T_{\mu\nu}^{\mu_1\mu_2} &=& q^2 g^{\mu_1}_\mu g^{\mu_2}_{\nu} -(g^{\mu_1}_\mu q_\nu+g^{\mu_1}_\nu q_\mu)q^{\mu_2} +g_{\mu\nu}q^{\mu_1}q^{\mu_2}
\eea
for $n=2,k=l=0$ for example, we find e.g.
  \bea
Y_{\mu\nu}^{T=4,n=2} &=& - \frac{4g}{q^6}\>T^{\mu_1\mu_2}_{\mu\nu} \>\mathcal{O}^{2(0,0)}_{n=2,\mu_1\mu_2} 
\eea
and
\bea
X_{\mu\nu}^{T=4,n=2} &=& -\frac{g}{2 q^6} \left [ \frac{q^\mu q^\nu}{q^2}-g^{\mu\nu}\right ]\left\{2\>q\cdot\mathcal{O}^{7(0)}_{n=2}-3\> q\cdot\mathcal{O}^{3(0)}_{n=2}-3\>q\cdot\mathcal{O}^{3(1)}_{n=2}\right\} \nn \\
&-&\frac{g}{2 q^6}\left[  g^{\mu\nu} - \frac{p^\mu q^\nu+p^\nu q^\mu}{p \cdot q} + \frac{q^2 p^\mu p^\nu}{(p\cdot q)^2} \right]\left\{\frac{1}{2}\>q\cdot\mathcal{O}^{3(0)}_{n=2}+\frac{1}{2}\>q\cdot\mathcal{O}^{3(1)} + 5\>q\cdot\mathcal{O}^{7(0)}_{n=2}\right\} \nn \\
\eea
Using the parametrization in Eq.(\ref{me}) reproduces the expression given in section \ref{wcs}. 
%%%%%%%%%%%%%%%%%%%%%%%%%%%%%%%%%%%%%%%
%												                            %
%				APPENDIX C		 				                            %
%											                                     %
%%%%%%%%%%%%%%%%%%%%%%%%%%%%%%%%%%%%%%%
\section{Operator Reduction}

Organizing the pole structure of the one loop corrections to the 2Q operators is a non-trivial task due to the appearance of various non-canoical, gauge-invariant operators appearing at the one loop level.  The following list of identities was used repeatedly to remove each non-canonical operator in favor of canonical ones.
\begin{small}
\begin{align*}
&\textbf{1.}& \gamma^\alpha \gamma^\beta \gamma^\chi & \ \ \ \ \ \ \ \ =&\gamma^\chi g^{\alpha\beta}+\gamma^\alpha g^{\beta\chi} -\gamma^{\beta}g^{\alpha\chi} + i \gamma^\sigma \gamma^5 \ep{\alpha}{\beta}{\chi}{\sigma}\\
&\textbf{2.}& (\ep{\alpha}{\beta}{\mu}{\lambda}\gamma^\lambda\gamma^5\mathcal{\overrightarrow{D}}_\beta) \psi&\ \ \ \ \ \ \ \ =&-i(D^\mu\gamma^\alpha-D^\alpha\gamma^\mu)\psi\\
&\textbf{3.}& \bar{\psi} (\ep{\alpha}{\beta}{\mu}{\lambda} \gamma^\lambda\gamma^5\mathcal{\overleftarrow{D}}_\alpha) &\ \ \ \ \ \ \ \ =&-i\bar{\psi}(D^\mu\gamma^\beta-D^\beta\gamma^\mu)\\
&\textbf{4.}&D^2&\ \ \ \ \ \ \ \ =&\sld{D}\sld{D}-\frac{1}{4}[\gamma^\alpha,\gamma^\beta][D_\alpha,D_\beta]\\
&\textbf{5.}& D_\mu F^{\mu\nu}&\ \ \ \ \ \ \ \ =&\frac{1}{2ig}([D_\mu,[D_\mu,D_\nu]]+D^2 D_\nu - D_\nu D^2 )\\
&\textbf{6.}&[D_\alpha,D_\beta]&\ \ \ \ \ \ \ \ =&igF_{\alpha\beta}\\
&\textbf{7.}&\slashed{D}\psi&\ \ \ \ \ \ \ \ =&0\\
&\textbf{8.}&[D_\beta,F^{\beta\rho}]&\ \ \ \ \ \ \ \ =&g \tau^a \> \sum_f \bar{\psi}_f\gamma^\rho \tau^a \psi_f
\end{align*}
\end{small}
As a simple example, one such non-canonical operator encountered in the one loop analysis of $Q_7^{(0,0)}$ is $\triangle^\mu \triangle^\nu \> \bar{\psi} \gamma_\mu \overleftarrow{D}_{\nu} D^2\psi$.  Applying identity \textbf{4.} and identity \textbf{1.} we find this operator is easily expressed in terms of a linear combination of canonical operators

\begin{align*}
=&\bar{\psi}\sld{\triangle}\triangle\cdot\overleftarrow{D}(\sld{D}\sld{D}-\frac{1}{4}[\gamma^\alpha,\gamma^\beta][D_\alpha,D_\beta)\psi\\
=&-\frac{1}{4}\bar{\psi}\sld{\triangle}\triangle\cdot\overleftarrow{D}[\gamma^\alpha,\gamma^\beta][D_\alpha,D_\beta]\psi\\
=&-\frac{1}{4}\triangle_\rho\bar{\psi}\gamma^\rho\triangle\cdot\overleftarrow{D}[\gamma^\alpha,\gamma^\beta]F_{\alpha\beta}\psi\\
=&-\frac{1}{4}\triangle_\rho\bar{\psi}\triangle\cdot\overleftarrow{D}(2i \gamma^\sigma \gamma^5 \ep{\alpha}{\beta}{\rho}{\sigma}+2\gamma^\beta g^{\alpha\beta}-2\gamma^\alpha g^{\beta\rho})F_{\alpha\beta}\psi\\
=&-\frac{1}{2}\triangle_\rho\bar{\psi}(\triangle\cdot\overleftarrow{D}(i \gamma^\sigma \gamma^5) \tilde{F}_{\rho\sigma})\psi-\frac{1}{2}\bar{\psi}(\triangle\cdot\overleftarrow{D}\sld{f})\psi-\frac{1}{2}\bar{\psi}(\triangle\cdot\overleftarrow{D}\sld{f})\psi\\
=&-\frac{1}{2}\bar{\psi}(\overleftarrow{d} \tilde{\sld{f}})\psi+i\bar{\psi}(\overleftarrow{d}\sld{f})\psi\\
=&-\frac{1}{2}Q^7_{k=1}+Q^8_{k=1}
\end{align*}

An exhaustive proof that all non-canonical, gauge invariant operators can be reduced to canonical form can be found in \cite{Jaffe:1982pm}.

%%%%%%%%%%%%%%%%%%%%%%%%%%%%%%%%%%%%%%%
%												                            %
%				APPENDIX D		 				                            %
%											                                     %
%%%%%%%%%%%%%%%%%%%%%%%%%%%%%%%%%%%%%%%
\section{Renormalization Group Running}
 \noindent The $\mu$-dependence of the coefficient functions on $Q^2$ is given by the solution to the RG equation
 \bea
\mu \frac{d}{d\mu} C^{j}_{n\tau} = \gamma_n^{ij} C^{i}_{n\tau}
 \eea 
the standard solution is given by
\bea
C_{n\tau}^{i}(Q^2/\mu^2,g)\simeq \sum_j\>C_{n\tau}^{j}(1,\bar{g}(t'))\>\text{T}\left[\text{exp}\left\{-\int_0^t\>\text{d}t'\>\gamma_{n\tau}(\bar{g}(t'))\right\}\right]_{ji}
\eea
And for a one-loop analysis, the T-ordering can be dropped when evaluating the integral.  Using the operator rescaling outlined in section \ref{counting} in the strong coupling, we may write $\gamma_{ji}(\bar{g}(t)) = \bar{g}^2(t) d_{ji}$, and dropping the subscript $n,\tau$ for simplicity
\bea
C_{i}\left(\frac{Q^2}{\Lambda^2},g,m\right) &=& \sum_j \> C_{j}\left(1,\bar{g}(0)\right)\>  R_{jm} \> \left\{\frac{1}{\beta_0 g^2} \> \text{Log}\left(\frac{Q^2}{\Lambda^2}\right)\right\}^{-\frac{d_{ml}}{2\beta_0}} \> R_{li}^{-1}\nn \\
\eea
where $R$ is a rotation matrix that diagonalizes the anomalous dimension, and we have made use of the following relations
\bea
\mu &=& \Lambda \>\text{exp}\left(\frac{1}{2\beta_0 g^2}\right) \nn \\
\bar{g}^2(t) &=& \frac{g^2}{1+2\beta_0\> g^2 \>t} \nn \\
\beta_0 &=& \frac{1}{(4\pi)^2}\left(\frac{11}{3}C_A - \frac{4}{3}T_f \>n_f\right) \nn \\
t &=& \frac{1}{2} \> \text{Log} \left(\frac{Q^2}{\mu^2}\right)
\eea
%%%%%%%%%%%%%%%%%%%%%%%%%%%%%%%%%%%%%%%
%												                            %
%				APPENDIX E - MATRIX ELMS    		                            %
%											                                     %
%%%%%%%%%%%%%%%%%%%%%%%%%%%%%%%%%%%%%%%
\section{Matrix Elements}\label{appendixE}
In this section we list the values of all matrix elements used to predict $M^{I=1}_{n=2,\tau=4}(Q^2)$.  In Table~\ref{rme} and \ref{totme} we include a brief description of the assumptions and denote the flavor structure of each operator for clarity.  Following the notation at the end of section~\ref{matrixelements}, we list the reduced matrix elements ($\mathcal{A}_{\pm}$) the four-quark operators, 
\bea\label{notation}
\mathcal{A}^{j,c}_{\pm} = \mathcal{A}^{j,2} \pm 2 \mathcal{A}^{j,4} + \mathcal{A}^{j,6}\\
\mathcal{A}^{j}_{\pm} = \mathcal{A}^{j,1} \pm 2 \mathcal{A}^{j,3} + \mathcal{A}^{j,5}
\eea
where the index $j$ denotes the flavor representation and $c$ denotes the presence of an SU(3) color generator.  
\begin{table}
\caption[]{Reduced matrix elements for the four-quark operators of each flavor representation.  All values have been computed at the input scale $5 \> \text{GeV}^2$ and the third column summarizes the methods used to compute the final value.}
\begin{center}
\begin{ruledtabular}
\begin{tabular}{ccc}
$j=A$,  (I=1,27) - $\bar{\psi}Q\psi\>\bar{\psi}Q\psi$ & Value (GeV$^2$)  & Method \\
\hline
$\mathcal{A}_+^{c} $     &  $5.2\times10^{-3}$  & Lattice~\cite{Gockeler:2001xw} \\
$\mathcal{A}_-^{c}  $    & $-10.4\times 10^{-4}$  &   Lattice~\cite{Gockeler:2001xw} \\
$\mathcal{A}_+  $   & $9.8\times 10^{-3}$ &   Lattice~\cite{Gockeler:2001xw} \\
$\mathcal{A}_-  $   &  $0.0$  & Lattice~\cite{Gockeler:2001xw}\\
\\
\hline
$j=A$, (I=1,$8$) - $\bar{\psi}Q\psi\>\bar{\psi}Q\psi$& Value (GeV$^2$)  & Method \\
\hline
$\mathcal{A}_+^{c} $     &  $5.2\times10^{-3}$  & $27\equiv 8A$ Assumption \\
$\mathcal{A}_-^{c}  $    & $-10.4\times 10^{-4}$  & $27\equiv 8A$ Assumption \\
$\mathcal{A}_+  $   & $9.8\times 10^{-3}$ &  $27\equiv 8A$ Assumption\\
$\mathcal{A}_-  $   &  $0.0$  & $27\equiv 8A$ Assumption\\
\\
\hline
$j=B$,  (I=1,$8$) - $\bar{\psi}Q^2\psi\>\bar{\psi}\psi$& Value (GeV$^2$)  & Method \\
\hline
$\mathcal{A}_+^{c} $     &  $-0.06\le \mathcal{A}_+^c \le 0.04$ & Phenomenology~\cite{Choi:1993cu} \\
$\mathcal{A}_+  $    & $-0.06\le \mathcal{A}_+^{c} \le 0.04$  & Assumption\\
\\
\hline
$j=C$, (I=1,$8$) - $\bar{\psi}Q^2\psi$& Value (GeV$^2$)  & Method \\
\hline
$\mathcal{A}^{j=C} $     &  $-0.1\le \mathcal{A}^{j=C} \le -0.04$ & Phenomenology~\cite{Choi:1993cu}
\\
%\hline
\end{tabular}
\label{rme}
\end{ruledtabular}
\end{center}
\end{table}

To make contact with the operators listed in Eq.(\ref{jaffe-basis-1}), we have written the reduced matrix elements using the notation $\langle Q^j \rangle$, where the index $j$ denotes basis operator of type $j$ appearing in Eq.(\ref{jaffe-basis-1}).  For all operators appearing in Table~\ref{totme}, we have set $n=2$ and specify the $k$ values for operators $Q^{7,8}$.
\begin{table}
\caption[]{All values of the reduced matrix elements used in computing $M^{I=1}_{n=2,\tau=4}\left(Q^2\right)$ at an input scale of $Q_0^2 = 5\>\text{GeV}^2$.  The methods and assumptions used to arrive at these values are summarized in the third column.}
\begin{center}
\begin{ruledtabular}
\begin{tabular}{ccc}
(I=1,27)  $\bar{\psi}Q\psi \>\bar{\psi}Q\psi$& Value (GeV$^2$)  & Method \\
\hline
$\langle Q^1 \rangle  $     &  $2.5\times10^{-3}$  & Lattice~\cite{Gockeler:2001xw}, \& Eq.(\ref{consts}) \\

$\langle Q^2 \rangle   $    & $-1.3\times 10^{-4}$  &   Lattice~\cite{Gockeler:2001xw}, \& Eq.(\ref{consts})  \\

$\langle Q^3 \rangle   $   & $2.5\times 10^{-3}$ &   Lattice~\cite{Gockeler:2001xw}, \& Eq.(\ref{consts})  \\

$\langle Q^4 \rangle   $   &  $-1.0\times 10^{-3}$  & Lattice~\cite{Gockeler:2001xw}, \& Eq.(\ref{consts})  \\

$\langle Q^5 \rangle   $   & $2.5 \times 10^{-3}$     & Lattice~\cite{Gockeler:2001xw}, \& Eq.(\ref{consts})  \\

$\langle Q^6 \rangle   $   & $-1.8\times 10^{-3}$     &  Lattice~\cite{Gockeler:2001xw}, \& Eq.(\ref{consts})  \\
%\hline
\\
\hline
(I=1, $8A$)   $\bar{\psi}Q\psi \>\bar{\psi}Q\psi$ & Value (GeV$^2$) & Method \\
\hline
$\langle Q^1 \rangle  $     &  $2.5\times10^{-3}$  & 27 $\equiv$ $8A$, Assumption\\

$\langle Q^2 \rangle   $    & $-1.3\times 10^{-4}$  &  27 $\equiv$ $8A$, Assumption \\

$\langle Q^3 \rangle   $   & $2.5\times 10^{-3}$ &   27 $\equiv$ $8A$, Assumption \\

$\langle Q^4 \rangle   $   &  $-1.0\times 10^{-3}$  & 27 $\equiv$ $8A$, Assumption \\

$\langle Q^5 \rangle   $   & $2.5 \times 10^{-3}$     & 27 $\equiv$ $8A$, Assumption \\

$\langle Q^6 \rangle   $   & $-1.8\times 10^{-3}$     & 27 $\equiv$ $8A$, Assumption \\
%\hline
\\
\hline
(I=1, $8B$)   $\bar{\psi}Q^2\psi \>\bar{\psi}\psi$ & Value (GeV$^2$) & Method \\
\hline
$\langle Q^1 \rangle  $     &  $-0.0015\le \langle Q^1\rangle \le 0.01 $  &  Phenomenology~\cite{Choi:1993cu}\\

$\langle Q^2 \rangle   $    & $-0.0015\le \langle Q^2\rangle \le 0.01 $ &   Phenomenology~\cite{Choi:1993cu} \\

$\langle Q^3 \rangle   $   & $-0.0015\le \langle Q^3\rangle \le 0.01 $ &   Phenomenology~\cite{Choi:1993cu} \\

$\langle Q^4 \rangle   $   & $-0.0015\le \langle Q^4\rangle \le 0.01 $  & Phenomenology~\cite{Choi:1993cu} \\

$\langle Q^5 \rangle   $   & $-0.0015\le \langle Q^5\rangle \le 0.01 $  & Phenomenology~\cite{Choi:1993cu} \\

$\langle Q^6 \rangle   $   & $-0.0015\le \langle Q^6\rangle \le 0.01 $  & Phenomenology~\cite{Choi:1993cu} \\
%\hline
\\
\hline
(I=1, $8C$)   $\bar{\psi}Q^2\psi $ & Value (GeV$^2$) & Method \\
\hline
$\langle Q^{7,(k=0)} \rangle  $     &  $-0.05 \le \langle Q^{7,k=0} \rangle \le -0.02$  &  Phenomenology~\cite{Choi:1993cu}, \& Eq.(\ref{2qconst})\\

$\langle Q^{7,(k=1)} \rangle   $    & $ 0.02 \le \langle Q^{7,k=0} \rangle \le 0.05$  &   Phenomenology~\cite{Choi:1993cu}, \& Eq.(\ref{2qconst}) \\

$\langle Q^{8,(k=0)} \rangle   $   & $-0.05 \le \langle Q^{7,k=0} \rangle \le -0.02$ &   $\langle Q^8 \rangle = \langle Q^7 \rangle$, Assumption \\

$\langle Q^{8,(k=1)}\rangle   $   &  $0.02 \le \langle Q^{7,k=0} \rangle \le 0.05$  & $\langle Q^8 \rangle = \langle Q^7 \rangle$, Assumption \\
\end{tabular}
\label{totme}
\end{ruledtabular}
\end{center}
\end{table}

\newpage

%%%%%%%%%%%%%%%%%%%%%%%%%%%%%%%%%%%%%%%
%												                            %
%				CONVENTIONS F.RULE				                            %
%											                                     %
%%%%%%%%%%%%%%%%%%%%%%%%%%%%%%%%%%%%%%%
\section{Conventions and Feynman Rules }
In this section, as elsewhere in the paper, we contract all free Lorentz indicies with a light light vector $\triangle_{\mu}$, where $\triangle^2 =0$ to project out the symmetrized and traceless portion of each operator.  Following \cite{Jaffe:1982pm} we write
\bea
f^\alpha &=& \triangle_\beta F^{\beta\alpha}\nn \\
\slashed{f} &=&\triangle_\beta\gamma_\alpha F^{\beta\alpha}\nn \\
d &=& i\triangle\cdot D \nn \\
\eea
where
\bea
 F^{a}_{\mu\nu} &=& \partial_{\mu}A_{\nu}^a - \partial_{\nu}A_{\mu}^a-i g f^{abc}A^{b}_\mu A^{c}_\nu \nn \\
 D_{\mu} &=& \partial_\mu + ig \tau^a A^a_\mu
\eea
And the QCD vertices and propagators are
%%%%%%%%
\begin{small}
\begin{equation}
  \label{eq:QCD3}
  \parbox{40mm}{
    \begin{center}
      \includegraphics{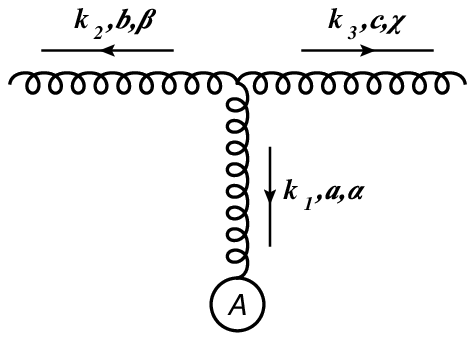}
    \end{center}
  }
\hspace{0.5 in}=-gf^{abc}\left\{g_{\alpha\chi} (k_\beta^1-k_\beta^3-\frac{1}{\xi}k_\beta^2)+g_{\alpha\beta}(k_\chi^2-k_\chi^1+\frac{1}{\xi}k_\chi^3)+g_{\chi\beta}(k^3_\alpha-k^2_\alpha)\right\}
\end{equation}
\end{small}
%%%%%%%%

%%%%%%%%
\begin{small}
\begin{equation}
  \label{eq:QV}
  \parbox{40mm}{
    \begin{center}
      \hspace{-7.0 in}\includegraphics{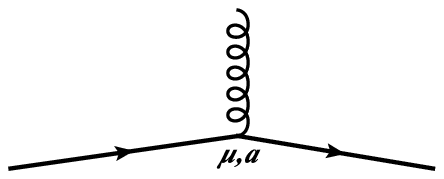}
    \end{center}
  }
\hspace{-3.0 in}=-i g \gamma^\mu \tau^a
\end{equation}
\end{small}
%%%%%%%%

%%%%%%%%
\begin{small}
\begin{equation}
  \label{eq:Prop}
  \parbox{40mm}{
    \begin{center}
      \hspace{-4.2 in}\includegraphics{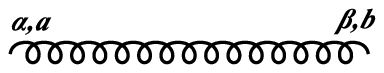}
    \end{center}
  }
\hspace{-1.6 in}= \frac{-i \delta_{ab}}{k^2}\left\{g^{\alpha\beta}-\left(1-\xi \right)\frac{k^\alpha k^\beta}{k^2}\right\}
\end{equation}
\end{small}
%%%%%%%%

%%%%%%%%
\begin{small}
\begin{equation}
  \label{eq:Prop}
  \parbox{40mm}{
    \begin{center}
      \hspace{-6.2 in}\includegraphics{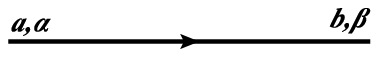}
    \end{center}
  }
\hspace{-2.6 in}= \frac{i (\slashed{k}+m_f)}{k^2-m_f^2+i \epsilon}
\end{equation}
\end{small}
%%%%%%%%
Where, in our calculations, we take $m_f=0$, and choose $\xi=1$.  The Feynman rules for the single gluon, 2Q operators
%%%%%%%%
\begin{small}
\begin{equation}
  \label{eq:Ops}
  \parbox{40mm}{
    \begin{center}
      \hspace{-2.4 in}\includegraphics{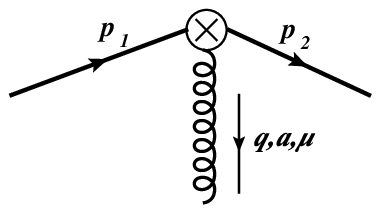}
    \end{center}
  }
\hspace{-0.6 in}\rightarrow \begin{cases} Q^{7}_{k=0} =& 2i\triangle_\alpha\tau^a\>\epsilon^{\alpha\beta\nu\mu}\>\gamma^\beta\gamma^5\>q_\nu\triangle\cdot p_1
\\
Q^7_{k=1} =&  -2i\triangle_\alpha\tau^a\>\epsilon^{\alpha\beta\nu\mu}\>\gamma^\beta\gamma^5\>q_\nu\triangle\cdot p_2
\end{cases}
\end{equation}
\end{small}
%%%%%%%%

%%%%%%%%
\begin{small}
\begin{equation}
  \label{eq:Ops}
  \parbox{40mm}{
    \begin{center}
      \hspace{-2.4 in}\includegraphics{TwoQOps.eps}
    \end{center}
  }
\hspace{-0.6 in}\rightarrow \begin{cases} Q^{8}_{k=0} =& -(\triangle\cdot q\> \gamma_\mu-\slashed{q}\triangle_\mu)\>\tau^a\triangle\cdot p_1 \\
Q^8_{k=1} =& (\triangle\cdot q\> \gamma_\mu-\slashed{q}\triangle_\mu)\>\tau^a\triangle\cdot p_2 
\end{cases}
\end{equation}
\end{small}

Feynman rules for the two-gluon, 2Q operators

%%%%%%%%
\begin{small}
\begin{equation}
  \label{eq:Ops}
  \parbox{40mm}{
    \begin{center}
      \hspace{0.0 in}\includegraphics{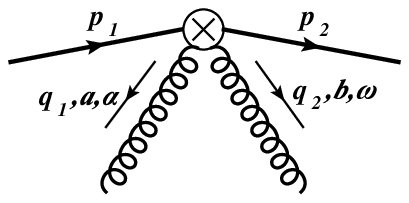}
    \end{center}
  }
\hspace{0.1 in}\rightarrow \begin{cases} Q^{7}_{k=0} =&ig\>\gamma_\beta\gamma_5\>\epsilon^{\mu\beta\alpha\nu}\>\Delta_\mu\left\{2\triangle_{\omega}\tau^a\tau^b\> q^1_\nu+i\tau^cf^{cab} g_{\omega\nu}\triangle\cdot p_1\right\} +\\
&ig\>\gamma_\beta\gamma_5\>\epsilon^{\mu\beta\omega\nu}\>\Delta_\mu\left\{2\triangle_{\alpha}\tau^b\tau^a\> q^2_\nu+i\tau^cf^{cba} g_{\alpha\nu}\triangle\cdot p_1\right\} \\
\\
Q^7_{k=1} =&-ig\>\gamma_\beta\gamma_5\>\epsilon^{\mu\beta\alpha\nu}\>\Delta_\mu\left\{2\triangle_{\omega}\tau^b\tau^a\> q^1_\nu+i\tau^cf^{cba} g_{\omega\nu}\triangle\cdot p_2\right\} -\\
&\>\>\>ig\>\gamma_\beta\gamma_5\>\epsilon^{\mu\beta\omega\nu}\>\Delta_\mu\left\{2\triangle_{\alpha}\tau^a\tau^b\> q^2_\nu+i\tau^cf^{cab} g_{\alpha\nu}\triangle\cdot p_2\right\} \\
\end{cases}
\end{equation}
\end{small}
%%%%%%%%

%%%%%%%%
\begin{small}
\begin{equation}
  \label{eq:Ops}
  \parbox{40mm}{
    \begin{center}
      \hspace{0.0 in}\includegraphics{TwoGlueOps.eps}
    \end{center}
  }
\hspace{0.1 in}\rightarrow \begin{cases} Q^{8}_{k=0} =&g\left\{(\tau^a\tau^b\triangle^\omega)(\triangle\cdot q^1 \gamma^\alpha-\slashed{q}^1\triangle^\alpha)-(ig \tau^cf^{cab})(\triangle^\alpha \gamma^\omega \triangle\cdot p_1)\right\} +\\
&g\left\{(\tau^b\tau^a\triangle^\alpha)(\triangle\cdot q^2 \gamma^\omega-\slashed{q}^2\triangle^\omega)-(ig \tau^cf^{cba})(\triangle^\omega \gamma^\alpha \triangle\cdot p_2)\right\} \\
\\
Q^8_{k=1} =&g\left\{-(\tau^a\tau^b\triangle^\alpha)(\triangle\cdot q^2 \gamma^\omega-\slashed{q}^2\triangle^\omega)+(ig \tau^cf^{cab})(\triangle^\alpha \gamma^\omega \triangle\cdot p_2)\right\} +\\
&g\left\{-(\tau^b\tau^a\triangle^\omega)(\triangle\cdot q^1 \gamma^\alpha-\slashed{q}^1\triangle^\alpha)+(ig \tau^cf^{cba})(\triangle^\omega \gamma^\alpha \triangle\cdot p_2)\right\} \\
\end{cases}
\end{equation}
\end{small}
%%%%%%%%
\bibliographystyle{h-physrev3.bst}
\bibliography{Higgs.bib}

\end{document}